\documentclass[
preprint,            % Format:   preprint, twocolum
showpacs,            % Pacs  :   showpacs, noshowpacs
preprintnumbers,     % Preprint: preprintnumbers, nopreprintnumbers
aps,                 % Society:
prd,                 % Style :   pra,prb,prc,prd,pre,prl,prstab,rmp
a4paper,             % Size  :   a4paper, ...
superscriptaddress,  % Title :   groupedaddress, superscriptaddress, unsortedaddress
unsortedaddress,
nofootinbib,         % Foonote:  footinbib, nofootinbib (in bibliography or in text)
tightenlines,        % Remove additional spaces in a line
floats               % Floating pictures and tables
]{revtex4}

\usepackage{graphicx}
\usepackage{color}
\usepackage{amssymb,latexsym}
\usepackage{varioref,xr-hyper}
\usepackage[draft=false,verbose=false,debug=false]{hyperref}

\newcommand{\lsim}{\,\raise 0.4ex\hbox{$<$}\kern -0.8em\lower 0.62ex\hbox{$\sim$}\,}
\newcommand{\gsim}{\,\raise 0.4ex\hbox{$>$}\kern -0.7em\lower 0.62ex\hbox{$\sim$}\,}
\newcommand{\arctg}{{\rm arctg}}
\newcommand{\Tcos}{{\rm Tcos}}
\newcommand{\Tsin}{{\rm Tsin}}
\newcommand{\si}{\sigma}

\newcommand{\be}{\begin{equation}}
\newcommand{\ee}{\end{equation}}
\newcommand{\bea}{\begin{eqnarray}}
\newcommand{\eea}{\end{eqnarray}}
\newcommand{\geneva}{D\'epartement de Physique Th\'eorique,
                     Universit\'e de Gen\`eve, 24 quai Ernest Ansermet,
                     CH--1211 Gen\`eve 4, Switzerland}
\newcommand{\sussex}{Centre for Theoretical Physics, University of Sussex,
                     Falmer, Brighton BN1 9QJ, United Kingdom}

\begin{document}

\title{Cosmological perturbations and the transition
       from contraction to expansion}

\author{\firstname{Cyril} \surname{Cartier}}
\email[E-mail address: ]{cyril.cartier@physics.unige.ch}
\affiliation{\geneva}

\author{\firstname{Ruth} \surname{Durrer}}
\email[E-mail address: ]{ruth.durrer@physics.unige.ch}
\affiliation{\geneva}

\author{\firstname{Edmund J.} \surname{Copeland}}
\email[E-mail address: ]{e.j.copeland@sussex.ac.uk}
\affiliation{\sussex}

\date{January 24, 2003}
\preprint{hep-th/0301198}

%======================================%
%<<<<<<<<<<<<< ABSTRACT >>>>>>>>>>>>>>>%
%======================================%

\begin{abstract}
We investigate both analytically and numerically the evolution of
scalar perturbations generated in models which exhibit a {\em
smooth} transition from a contracting to an expanding Friedmann
universe. If the perturbation equations are formulated as second
order equations for either the Bardeen potential $\Psi$ or the
curvature perturbation on uniform comoving hypersurfaces $\zeta$,
at best one of them can stay regular during the transition. We
find that the resulting spectral index in the late radiation
dominated universe {\em depends} on which of these two variables
passes regularly through the transition. The results can be
parametrized by the exponent $q$ defining the rate of contraction
of the universe, or equivalently through the equation of state
$w=(2-q)/3q$ of the background fluid. For $q \geqslant
-\frac{1}{2}$ we find that there are {\em no} stable cases where
both $\Psi$ and $\zeta$ are regular during the transition. In
particular, for $0<q\ll 1$, we find that the resulting spectral
index is close to scale invariant if $\Psi$ is regular, whereas it
has a steep blue behavior if $\zeta$ is regular. We also show that
as long as $q\leqslant 1$, perturbations remain small during
contraction in the sense that there exists a gauge in which all
the metric and matter perturbation variables are small. This work
has important implications for the current debate concerning the
nature of perturbations evolving through a collapsing regime into
an expanding one: it shows that if in the ekpyrotic model, where
$0<q\ll 1$, the Bardeen potential passes regularly through the
transition, this leads to a nearly scale invariant spectrum with
$n=1-2q$, whereas in the case of dilaton-driven string cosmology
we have the opposite situation. There it is assumed that $\zeta$
passes regularly through the transition, leading to a very blue
spectrum of highly suppressed perturbations.
\end{abstract}

\maketitle

\section{Introduction}
\label{s:introduction}

Recently, it has been argued that a slowly contracting universe
which transits smoothly over to an expanding radiation dominated
era may lead to a scale invariant spectrum of adiabatic density
fluctuations~\cite{Khoury:2001zk,Durrer:2002jn,Peter:2002cn,Khoury:2001wf}.
There has been quite an intense debate on the question of whether
the resulting spectrum after the transition to the radiation era
indeed is scale invariant or whether it has a steep blue spectrum
with $n\simeq 3$. The latter result has been advocated mainly
by~\cite{Brandenberger:2001bs,Lyth:2001pf,Hwang:2001ga}. Gratton
\textit{et al.} have now gone further arguing that the ekpyrotic
or cyclic scenario is the only robust case where a scale-invariant
spectra can be found (along with inflation for the case of an
expanding universe) \cite{Gratton:2003pe}.

If the equations of motion governing the transition for the
background and the perturbations were known, this problem could be
solved by integrating them numerically. However it is likely that
the evolution in this high curvature regime will contain full
string theory and we do not even know whether the variables of the
low energy theory are appropriate for the description of this
regime. One possibility which has been studied in the past is the
inclusion of first order corrections in the string scale $\alpha'$
and/or the coupling constant $g_s$ (see, e.g.
\cite{Antoniadis:1994jc,Brustein:1998cv,Foffa:1999dv,Cartier:1999vk}).
In this context it has been shown that, within a certain range of
coefficients for the terms added to the tree level Lagrangian, one
can exit from the high curvature regime and enter into a radiation
dominated phase~\cite{Cartier:1999vk}. In~\cite{Cartier:2001is}
the corrections of the perturbation equations have been derived
and have been solved for dilaton-driven string cosmology. Lately,
Tsujikawa \textit{et al.}~\cite{Tsujikawa:2002qc} have used these
equations to study the ekpyrotic
model~\cite{Khoury:2001wf,Khoury:2001zk}. Even though they have
followed the perturbations through a regularized transition, we
will argue that their method invariably leads to $n\simeq 3$ and
does not allow for a decision whether a blue, $n\simeq 3$ or a
scale-invariant, $n\simeq 1$ spectrum of perturbations is
obtained. In this paper we first study a general transition which
satisfies relatively mild criteria and we formulate conditions for
the transition which lead to either of the two spectral indices.
We can decide under which conditions either the spectral index
$n=1$ as advocated by Durrer and Vernizzi~\cite{Durrer:2002jn} or
$n=3$ put forward in~\cite{Finelli:2001sr} and others is obtained
in the ekpyrotic model. This model is a special case of the more
general class of contracting universes discussed here.

The rest of this paper is organized as follows: In
Sec.~\ref{s:background-perturbation} we present the aspects of
cosmological perturbation theory needed in this paper. We then
discuss the transition from a contracting to an expanding phase
emphasizing the differences between such a transition and that
associated with the transition between a conventional inflationary
phase to radiation. We also formulate the problem encountered when
inferring the perturbation spectrum after the transition from the
one before the transition. In Sec.~\ref{s:General solutions} we
study the behavior of perturbations during a transition and find
that the resulting spectral index depends on mutually excluding,
simple regularity conditions for the transition, which we
formulate in detail.

In Sec.~\ref{s:toy} we study numerical toy models for the
transition where we exemplify the general results obtained in the
previous section and analyze the stability of the numerically
obtained spectral indices. Then we formulate our regularity
condition as a theorem and we study the amplitude of
perturbations. In Sec.~\ref{s:discussion} we comment on the
findings in previous work and we summarize our results.

\section{Background and perturbations in simple
         contracting and expanding universes}
\label{s:background-perturbation}

In this section we repeat the basic equations for the Friedmann
background and adiabatic first order perturbations. We then
discuss initial conditions for perturbations in a contracting
universe and explain how a different spectral index is obtained
depending on the perturbation variable used for the transition.

\subsection{The background}
\label{ss:background}

We consider a Friedmann universe with negligible spatial
curvature, which is first contracting for $-\infty <\eta<-\eta_s$
and hence its (spacetime) curvature is growing. The variable
$\eta$ denotes conformal time and $\eta_s>0$ is the moment when we
enter a high-curvature regime. We call this period the
``pre-big-bang phase''. At $\eta \sim -\eta_s$ corrections coming
from some underlying theory, which reduces to general relativity
when the curvature is sufficiently small, become important and we
assume that they regularize the geometry and lead to an expanding
universe at $\eta>\eta_s$. We also assume that radiation is
produced during the transition so that for $\eta>\eta_s$ (the
``post-big-bang'' phase) the universe can be described as a
radiation dominated Friedmann universe. In the pre- and
post-big-bang phases, $|\eta|\gg \eta_s$, we have
\bea
 {\cal H}^2 &=& \frac{\kappa^2}{3} \rho a^2~, \label{e:Friedmann1}\\
 {\cal H}'  &=& - {\kappa^2 \over 6} (\rho+3 P) a^2
        = - {\cal H}^2\frac{1+3 w}{2} ~,  \label{e:Friedmann2}
\eea
where the equation of state of the background fluid is $w =
P/\rho$ and  $\kappa^2 = 8\pi G = 2/M_P^2$. $M_P = 1/\sqrt{4\pi G}
= 3.4\times 10^{18}$ GeV is the reduced Planck mass. ${\cal
H}=a'/a =\dot a = Ha$ is the comoving Hubble parameter. A prime
denotes a derivative with respect to conformal time $\eta$,
whereas a dot denotes a derivative with respect to physical time
$t$, defined by $dt=ad\eta$. We shall also use
\be \label{e:w'}
 w'=3(w-c_s^2)(1+w){\cal H}  \quad\mbox{ where }\quad
 c_s^2= \left(\frac{dP}{d\rho}\right)_{\scriptsize{\textrm{ad}}}
\ee
is the adiabatic sound speed (if $\rho \neq$ const, $c_s^2 =
P'/\rho'$). We will be especially interested in phases during
which $P/\rho=w=$ const. Then $c_s^2 =$ const $=w$ and the scale
factor evolves like a power law,
\be \label{e:evolution-a}
 a = \left|\frac{\eta}{\eta_s}\right|^q~, \qquad
 q = \frac{2}{1+3w}~.
\ee
An inflationary phase, defined by $1+3w<0$, is thus realized when
$q<0$. During the radiation dominated post-big-bang phase $w=1/3$,
hence $q=1$.

We will consider a scalar field dominated pre-big-bang phase.
During this phase we have
\be
 \rho = {\varphi'^2 \over 2a^2} +V(\varphi) ~,\qquad
 P = {\varphi'^2 \over 2a^2} - V(\varphi) ~,
\ee
so that
\be
 w+1 ={{\varphi'}^2\over 3{\cal H}^2} ~.
\ee
For a scalar field, $w$ is constant only for the following
possibilities:
\be
 w = \left\{ \begin{array}{lll}
      1       & \mbox{ if } & V=0 ~ ,\\
     -1       & \mbox{ if } & \varphi'=0  ~,\\
     \lambda^2/3-1  & \mbox{ if } & V=V_0\exp(-\lambda\varphi)~,\quad V_0\neq 0~,
     \end{array} \right.
\ee
and, correspondingly,
\be
  q = \left\{ \begin{array}{lll}
       1/2  & \mbox{ if } & V=0  ~,\\
       -1   & \mbox{ if } & \varphi'=0 ~ ,\\
       \frac{2}{\lambda^2-2} & \mbox{ if } & V=V_0\exp(-\lambda\varphi)~,
       \quad V_0\neq 0~.
       \end{array} \right.
\ee

\subsection{Perturbations}
\label{ss:perturbation-u-v}

We now discuss perturbation theory in a Friedmann universe
(neglecting spatial curvature, $K=0$) with a scalar field or a
perfect fluid, like radiation.

We first consider the linear perturbation equation for the Bardeen
potential (see, e.g.~\cite{Mukhanov:1992tc,Durrer:1993db}):
\bea
 \Psi'' + 3 {\cal H} (1+c_s^2) \Psi' +
 \left[ 2 {\cal H}' + (1+3 c_s^2) {\cal H}^2 - \Upsilon \Delta \right]
 \Psi = 0~.
 \label{e:Psi-eqn}
\eea
This equation is valid for adiabatic perturbations of a fluid,
with $\Upsilon = c_s^2$, or for a simple scalar field, with
$\Upsilon = 1$ (see, {\em e.g.}~\cite{Mukhanov:1992tc}).

If we define the canonical variable
\be \label{e:def-u-Psi}
 u = {M_{P}a\over \sqrt{{\cal H}^2-{\cal H}'}}\Psi ~,
\ee
$u$ satisfies the equation~\cite{Mukhanov:1992tc}
\be
 u'' +(\Upsilon k^2-\theta''/\theta)u =0~,  \label{e:u-eqn}
\ee
for
\be  \label{e:def-theta}
 \begin{array}{lcl}
 & \qquad & \theta = \theta_1 =
 {{\cal H} \over a\sqrt{{2\over 3}({\cal H}^2 -{\cal H}')}}~,\\
 \mbox{or} &&\\
 & \qquad &\theta = \theta_2 =  \theta_1\int{d\eta\over \theta_1^2}~.
\end{array}
\ee
If we restrict ourselves to the case $w=c_s^2=$ const, the mass
term in Eq.~({\ref{e:Psi-eqn}), namely, $2 {\cal H}' + (1+3 c_s^2)
{\cal H}^2$, vanishes by the use of the background Einstein
equations~(\ref{e:Friedmann1}) and (\ref{e:Friedmann2}). For these
backgrounds which have $a \propto |\eta|^q$, where $q$ is given in
Eq.~(\ref{e:evolution-a}), we find
\be  \label{e:Psi-eqn-reduced}
 \Psi''+ 2{\cal H}{1+q\over q} \Psi' + \Upsilon k^2 \Psi = 0 ~.
\ee
The $u$ equation then simply becomes a Bessel differential
equation,
\be \label{e:u-eqn-reduced}
 u''+\left(\Upsilon k^2-{q(q+1)\over\eta^2}\right)u =0~.
\ee
The correct normalization to the incoming vacuum at $\eta \to
-\infty$ determines the initial conditions such that (see,
e.g.~\cite{Durrer:2002jn})
\be  \label{e:u-initial-cond}
 u ={\sqrt{\pi |k\eta|}\over k^{3/2}}H_\mu^{(2)}(k\eta) ~,
\ee
with $\mu=q+1/2$. Here $H_\mu^{(2)}$ denotes the Hankel function
of the second kind and of order $\mu$. At ``late times'' when
$|k\eta| \ll 1$ but still $\eta\ll -\eta_s$ we may neglect the
$k^2$ term in Eq.~(\ref{e:u-eqn-reduced}) and find the
super-Hubble scale solution
\be \label{e:u-superHubble}
 u = A_-\left|\frac{\eta}{\eta_s}\right|^{-q} +
     B_-\left|\frac{\eta}{\eta_s}\right|^{1+q} = u_A + u_B  ~.
\ee
The coefficients $A_-$ and $B_-$ are determined by the initial
solution (\ref{e:u-initial-cond}),
\be \label{e:def-uAB-neg}
 A_- \simeq k^{-3/2}(k\eta_s)^{-q}~, \quad
 B_- \simeq k^{-3/2}(k\eta_s)^{1+q}~,
\ee
hence they have the spectra~\cite{Durrer:2002jn}
\bea \label{e:spectre-uAB-neg}
 P_{A_-} &=& |A_-|^2k^3  \simeq (k\eta_s)^{-2q}
 \propto k^{n_{A_-} -1}~,
 \quad~ n_{A_-} =1-2q~, \\
 P_{B_-} &=& |B_-|^2k^3 \simeq (k\eta_s)^{2+2q}
 \propto k^{n_{B_-} -1}~,
 \quad n_{B_-} =3+2q~.
\eea
Furthermore, $|u_B/u_A| \simeq |k\eta|^{1+2q}$. Hence for
$q>-1/2$, the $A$ mode, $u_A$, dominates at late time over the $B$
mode, $u_B$.

Another perturbation variable often used is the curvature
perturbation on uniform comoving hypersurfaces~\cite{Lyth:1985aa}
\be \label{e:def-zeta-Psi}
 \zeta = { {\cal H}\Psi' +{\cal H}^2\Psi \over {\cal H}^2-{\cal H}'} + \Psi~.
\ee
A simple substitution using Eq.~(\ref{e:Psi-eqn}) and the
background equations yields
\be \label{e:zeta-constant}
 \zeta' = -k^2{\Upsilon{\cal H}\over {\cal H}^2-{\cal H}'}\Psi ~,
\ee
hence on super-Hubble scales, $|k/{\cal H}|\ll 1$, this variable
is conserved. For ordinary inflationary models, it is therefore
usually sufficient to compute $\zeta$ at the time of the Hubble
radius crossing during inflation to obtain its value in the
radiation dominated era. Furthermore, since during radiation
$\zeta = (3/2) \Psi$, this simply gives the Bardeen potential.

The evolution of $\zeta$ is closely related to the canonical
variable $v$ defined by
\be \label{e:def-v}
 v = -{M_{P}a\sqrt{{\cal H}^2-{\cal H}'}\over \sqrt{\Upsilon}{\cal H}}\zeta~.
\ee
This variable satisfies the equation~\cite{Mukhanov:1992tc}
\be \label{e:v-eqn}
 v'' +(\Upsilon k^2- z''/z)v =0~,
\ee
where
\be  \label{e:def-z}
 \begin{array}{lcl}
 & \qquad & z = z_1 = {a\sqrt{{\cal H}^2-{\cal H}'}\over \sqrt{\Upsilon} {\cal H}} ~,\\
 \mbox{or} &&\\
 & \qquad & z_2 =  z_1\int{d\eta\over z_1^2}~.
\end{array}
\ee
Note that the relation between $v$ and $\zeta$ is
$v=-M_Pz_1\zeta$. Equation (\ref{e:v-eqn}) is invariant under the
``duality'' $z_1\to z_2$, in the same way as Eq.~(\ref{e:u-eqn})
is invariant under $\theta_1\to \theta_2$.

In~\cite{Mukhanov:1992tc} it is shown that $v$ appears in the
perturbed action as a canonical scalar variable. Hence on
sub-Hubble scales, $|k/{\cal H}|\gg 1$, it satisfies the initial
condition $v_{in} = \exp(-ik\eta)/\sqrt{k}$.

As before, we now concentrate on the case $w=c_s^2=$ const, with
the scale factor given in Eq.~(\ref{e:evolution-a}). Then
\be
  z_1 \propto a \quad \mbox{ and } \quad
  v = {\rm const}\times M_P a\zeta~,
\ee
and the constant of order unity depends on $q$ and $\Upsilon$.

During the pre-big-bang phase Eq.~(\ref{e:v-eqn}) then also
becomes a Bessel differential equation,
\be \label{e:v-eqn-reduced}
 v'' +\left(k^2- {q(q-1)\over \eta^2}\right)v =0~.
\ee
We have set $\Upsilon=1$ throughout the evolution, which should be
fine as we are considering mainly super-Hubble terms, which
satisfy $k^2\Upsilon \ll z''/z$ during the transition. We have
confirmed in numerical simulations that allowing $\Upsilon$ to
vary during the evolution does not affect the key results we
present which concern the spectral index.

The solution with the correct initial conditions is
\be  \label{e:v-initial}
 v ={\sqrt{\pi |k\eta|}\over k^{1/2}}H_\nu^{(2)}(k\eta) ~,
\ee
with $\nu=1/2-q$. At ``late times'' when $|k\eta| \ll 1$ but still
$\eta\ll -\eta_s$ we may neglect the $k^2$ term in
Eq.~(\ref{e:v-eqn-reduced}) and find the super-Hubble scale
solution
\be \label{e:v-superHubble}
 v = C_-\left|\frac{\eta}{\eta_s}\right|^{1-q} +
     D_-\left|\frac{\eta}{\eta_s}\right|^{q} = v_C + v_D  ~.
\ee
The coefficients $C_-$ and $D_-$ are determined by the initial
solution (\ref{e:v-initial}),
\be \label{e:def-vCD-neg}
 C_- \simeq (k\eta_s)^{\nu}\eta_s^{1/2}~, \quad D_-
 \simeq(k\eta_s)^{-\nu}\eta_s^{1/2}~.
\ee
The spectra obtained depend on the value of $q$. One
finds~\cite{Durrer:2002jn}
\bea
 P_{C_-} &=& |C_-|^2k^3
         \simeq (k\eta_s)^{4-2q}\eta_s^{-2} \propto k^{n_{C_-}-1}~,
         \quad\, n_{C_-} =5-2q~, \\
 P_{D_-} &=& |D_-|^2k^3
         \simeq (k\eta_s)^{2+2q}\eta_s^{-2} \propto k^{n_{D_-}-1}~,
         \quad n_{D_-} =3+2q~.\label{e:spectre-vCD-neg}
\eea
Here $|v_C/v_D|\simeq |k\eta|^{1-2q}$, hence the $D$ mode
dominates for $q<1/2$, while the $C$ mode dominates for $q>1/2$.
Finally we want to note that $u$ and $v$ are related via
\bea
 v &=& {- \theta_1\over \sqrt{\Upsilon}}(u/\theta_1)'~,
       \label{e:vu-relation}\\
 u &=& {z_1\over k^2\sqrt{\Upsilon}}(v/z_1)'~.
       \label{e:uv-relation}
\eea
It is interesting to note that from the lowest order
approximations for $u$ and $v$ given in
Eqs.~(\ref{e:u-superHubble}), (\ref{e:v-superHubble}) the
equivalences (\ref{e:vu-relation}) and (\ref{e:uv-relation}) of
$u$ and $v$ cannot be recovered. Only when we go to the next order
in the term proportional to $\theta_1$ or $z_1$, respectively (or
when using the full Bessel function solution) does the above
equivalence give $u_A \longleftrightarrow v_C$ and $u_B
\longleftrightarrow v_D$  along with
\be
 n_A+4=n_C~,\qquad n_B=n_D~.
\ee

\subsection{The problem of a transition from contraction to expansion}
\label{s:problem}

Let us first consider the case where  $q$ is in the interval
$-1/2\leqslant q \leqslant 1/2$. Even though $q<0$ does not
represent a contracting phase, no difference of the following
arguments arises from letting $q$ decrease until $-1/2$ (usual
inflation has $q\gsim -1$).

Comparing the amplitudes of the modes of $u$ and $v$, we see that
at the transition to the expansion phase, $|\eta|\sim \eta_s$, we
have $u_A\gg u_B$ and $v_D\gg v_C$ for cosmologically interesting
scales with $k\sim 1/\eta_0 \sim 10^{-30}/\eta_P$. Here $\eta_0$
denotes the value of conformal time today and $\eta_P = M_P^{-1}$.
Naively, we therefore expect that immediately {\em after the
transition}
\be \label{e:naive-uv-1}
 u \simeq A_- \quad\mbox{and}\quad v \simeq D_-
 \quad\mbox{ for }\quad
 -1/2\leqslant q \leqslant 1/2~.
\ee
Since in the radiation dominated era $u \propto \eta^2$ and $v \propto
\eta$ on super-Hubble scales, we expect during the radiation phase
\be \label{e:naive-uv-2}
 u \simeq A_-(\eta/\eta_s)^2 \quad\mbox{ and }\quad
 v \simeq D_-(\eta/\eta_s)   \quad\mbox{ for }\quad
 k\eta\ll 1 ~.
\ee
From the relations of $u$ and $\Psi$ as well as $v$ and $\zeta$
during the radiation dominated phase, this gives
\be \label{e:naive-Psi-zeta}
 \Psi \simeq \sqrt{2}~A_-/(\eta_sM_P) \quad\mbox{ and }\quad
 \zeta \simeq D_-/M_P  ~.
\ee

For $q\neq -1/2$, this naive result is clearly in contradiction
with the fact that during the radiation dominated era $\Psi$ and
$\zeta$ differ only by a constant since, according to
Eqs.~(\ref{e:spectre-uAB-neg}) and (\ref{e:spectre-vCD-neg}),
$\Psi$ would have the spectral index $n_\Psi= n_{A_-}=1-2q$ while
$\zeta$ would have the spectral index $n_\zeta= n_{D_-}=3+2q$.

For $q\leqslant -1/2$ the $B$ mode of $u$, $u_B$ dominates (for
$q=-1/2$, $u_A$ and $u_B$ are of the same order) and we expect
$\Psi$ to have the spectrum $n_\Psi=n_{B_-}=3+2q=n_{D_-}=n_\zeta$,
hence we obtain the same spectrum as $\zeta$ in the radiation era,
so that there is no contradiction.

For $q > 1/2$ the $C$ mode of $v$, $v_C$  dominates and hence
$\zeta$ actually has the spectrum $n_\zeta= n_{C_-} =5-2q$, which
is in even worse disagreement with the naively expected spectrum
for $\Psi$. This contradictory situation is shown on the right
hand panel of Fig.~\ref{f:indices_pbb}. Since for ordinary
inflation $q\sim -1$, this problem has never been realized when
studying usual inflation.

The simplest possibility which could resolve the issue is to note
that the decaying mode of $u$ ($q>-1/2$) during the pre-big-bang
phase, $u_B$, has the same spectrum as $v_D$. Hence if the
$u$-growing mode during the pre-big-bang phase is entirely
converted into the decaying mode after the transition and
therefore cannot be seen late in the post-big-bang era, we expect
the spectrum $n=3+2q$ in the radiation era. This argument has been
put forward in~\cite{Brandenberger:2001bs}, where the authors have
shown that this is exactly what happens if the transition is
defined by a vanishing jump in the metric and the second
fundamental form on the constant energy hypersurface. Similar
arguments have also been presented
in~\cite{Lyth:2001pf,Hwang:2001ga}. They led these authors to the
conclusion that the correct spectrum, evaluated sufficiently long
after the pre~$\to$~post transition so that the decaying mode has
died away, is $n=n_\Psi=n_\zeta=3+2q$. If this is correct the
spectrum of the ekpyrotic model is very blue and in contradiction
to the observed close-to-scale invariant spectrum. The same
argument for the original pre-big-bang model of
Veneziano~\cite{Veneziano:1991ek,Gasperini:2002bn}, where the
scalar field potential vanishes and hence $q=1/2$, led to the
conclusion that the dilaton perturbation spectrum is very blue
with $n=4$~\cite{Brustein:1995kn,Deruelle:1995kd}.

In~\cite{Durrer:2002jn} this argument has been criticized for two
main reasons. First, the background second fundamental form given
by ${\cal H}/a$ has to jump, even to change sign, in a transition
from contraction to expansion. It then seems quite unnatural to
require its perturbation to vanish. Second, if the matching
conditions are posed on an only slightly different hypersurface,
the naively expected spectral index, $n_\Psi = 1-2q$ is obtained.
This '`instability'' of the index $n_\Psi = 3+2q$ will also be
illustrated in Sec.~\ref{s:toy} with numerical studies of a simple
toy model.

The above argument cannot be used if $q>1/2$ because $u$ has no
mode with spectral index $5-2q$. In this case, agreement can only
be achieved if the dominant contribution to $v$, $v_D$, is also
transferred entirely into the decaying mode so that late after the
transition $v$ and hence $\zeta$ still have the spectral index
$n_{D_-}=3+2q$. However, this is not possible: As one easily
concludes, e.g. from~\cite{Brandenberger:2001bs} or
\cite{Lyth:2001pf},  a transition on the constant energy
hypersurface, where the growing mode of $u$ is transferred
completely into the decaying mode, preserves $\zeta$, hence
$\zeta$ has the same spectrum after the transition as before,
$n_\zeta=5-2q$ which does not agree with the spectrum of $\Psi$
which in this case is $n_\Psi = 3+2q$.

Hence if the transition is such that both $\Psi$ and $\zeta$
correspond after the transition to one of their modes before the
transition, the obtained spectral index must be $n=3+2q$. As we
have shown, this cannot happen for $q>1/2$ if the transition is
``simple'', i.e. does not modify the spectrum of either $\Psi$ or
$\zeta$.

If the spectral index after the transition is $n=1-2q$ as promoted
in~\cite{Durrer:2002jn} for $q>-1/2$, the variable $\zeta$ makes a
$k$-dependent jump at the transition. If $n=5-2q$ is obtained as
in~\cite{Finelli:2001sr}, $\Psi$ makes a $k$-dependent jump.
Furthermore, if $n=3+2q$ is obtained for $q>-1/2$, the growing
mode of $\Psi$ before the transition has to be converted entirely
into the decaying mode. For $q>1/2$ also this no longer helps
resolve the problem, and one of the two variables $\zeta$ or
$\Psi$ must be modified in a $k$-dependent way during the
transition.

\section{General solutions of the perturbation equations through the transition}
\label{s:General solutions}

Having explained the problem, but before discussing possible
resolutions, let us collect some generic facts about a transition
from contraction to expansion. Clearly, to have such a transition
${\cal H}$, ${\cal H}'$ and $\dot H = ({\cal H}/a)'/a = ({\cal
H}'-{\cal H}^2)/a^2$ have to change sign. Within the framework of
general relativity (neglecting spatial curvature) this requires
$\rho+P<0$ and therefore cannot be achieved with a scalar field
(with standard kinetic term). If a positive spatial curvature is
added, the scalar field initial condition can be fine tuned such
that close to the collapse the curvature term dominates over the
scalar field contributions, and a transition from contraction to
expansion can be achieved with a standard scalar
field~\cite{Page:1984}.

In this section we want to discuss the problem outlined above
without specifying any details of the transition. For this, we
first discuss the linear second order differential equation
\be
 x'' + \left(k^2 -V_x\right)x=0 ~, \qquad V_x = {s''\over s}~,
\label{e:x-eqn}
\ee
which we have encountered in the previous section. Here the
variables $(x,s)$ stand for either $(u,\theta)$ or $(v,z)$. The
factor $\Upsilon$ in front of the $k^2$ term is disregarded since
it is irrelevant for our considerations which mainly concern
super-Hubble scales. We notice that Eq.~(\ref{e:x-eqn}) is
invariant under the duality transformation $s=s_1 \to s_1 \int
d\eta/s_1^2 \equiv s_2$. If $s$ is a power law,
$s_1=\left|\eta/\eta_s\right|^\gamma$, we can set
$s_2=\left|\eta/\eta_s\right|^{1-\gamma}$. The duality property of
Eq.~(\ref{e:x-eqn}) has been discussed in \cite{Wands:1998yp}. If
$s$ and $1/s$ are bounded in the interval $[\eta_{in},\eta]$, so
that
\be \label{e:bound}
 \infty >C =\max_{\{\eta_{in}\leqslant\eta_1\leqslant\eta\}}
 \Big(s(\eta_1)^2,1/s(\eta_1)^2\Big)~,
\ee
this equation has the general
solution~\cite{Brustein:1998kq,Gasperini:2002bn}
\be \label{e:sol-Tsin-Tcos}
 x = s\Big[\alpha\Tcos(s,k)+\beta\Tsin(s,k)\Big]~,
\ee
where $\Tcos$ and $\Tsin$ are defined by
\bea
 \Tcos(s,k)
 &\equiv& 1 - k^2\int_{\eta_{in}}^\eta {d\eta_1\over s^2(\eta_1)}
          \int_{\eta_{in}}^{\eta_1} d\eta_2 s^2(\eta_2) \nonumber\\
 && \hspace{-0.2cm}
    + k^4\int_{\eta_{in}}^\eta {d\eta_1\over s^2(\eta_1)}
    \int_{\eta_{in}}^{\eta_1} d\eta_2 s^2(\eta_2)\int_{\eta_{in}}^{\eta_2}
    {d\eta_3\over s^2(\eta_3)} \int_{\eta_{in}}^{\eta_3} d\eta_4 s^2(\eta_4)
    - k^6\cdots \quad , \label{e:def-Tcos}\\
 \Tsin(s,k)
 &\equiv& k \int_{\eta_{in}}^\eta {d\eta_1\over s^2(\eta_1)} \nonumber \\
 && \hspace{-0.2cm}
    - k^3\int_{\eta_{in}}^\eta {d\eta_1\over s^2(\eta_1)}
    \int_{\eta_{in}}^{\eta_1} d\eta_2 s^2(\eta_2)\int_{\eta_{in}}^{\eta_2}
    {d\eta_3\over s^2(\eta_3)} + k^5\cdots  \quad . \label{e:def-Tsin}
\eea
When expressing a given solution in terms of $\Tcos$ and $\Tsin$
the coefficients $\alpha$ and $\beta$ will depend on the initial
value $\eta_{in}$ chosen. But as long as $s$ and $1/s$ are
bounded, the sums~(\ref{e:def-Tcos}) and (\ref{e:def-Tsin}) always
converge since the terms in this sum are bounded, e.g. by the
terms in the series expansion for
$\cos\left[Ck(\eta-\eta_{in})\right]$ and
$\sin\left[Ck(\eta-\eta_{in})\right]$, respectively. Here $C$ is
the bound from Eq.~(\ref{e:bound}) above.

To relate this solution with the results of Sec.~\ref{s:problem},
we choose $\eta_{in}$ such that $k|\eta_{in}|\ll 1$, but
$\eta_{in}\ll -\eta_s$. If $s$ obeys a simple power law,
$s=|\eta/\eta_s|^\gamma$, hence $s''/s=\gamma(\gamma-1)/\eta^2$,
we obtain to lowest order
\be
 \Tcos(s,k) = 1~,
\ee
and
\be
 \Tsin(s,k) = {k\eta_s\over 1-2\gamma}
 \left[\left|\frac{\eta_{in}}{\eta_s}\right|^{1-2\gamma}-
 \left|\frac{\eta}{\eta_s}\right|^{1-2\gamma}\right] ~,
\ee
so that
\be
x = \left[\alpha + {\beta k\eta_s\over 1-2\gamma}
    \left|\frac{\eta_{in}}{\eta_s}\right|^{1-2\gamma} \right]
    \left|\frac{\eta}{\eta_s}\right|^\gamma
    - \beta{k\eta_s\over 1-2\gamma}
    \left|\frac{\eta}{\eta_s}\right|^{1-\gamma}~.
\label{e:x-sol-neg}
\ee (For $\gamma=1/2$ the powers in the $\Tsin$ integral becomes a
logarithm, but we shall neglect this logarithmic correction here.)
There are several facts to note at this point:
\begin{itemize}
\item Only two of the three parameters $\eta_{in},\alpha,\beta$
which determine the initial conditions are independent.

\item As in Eqs.~(\ref{e:def-theta}), (\ref{e:def-z}), two pump
fields $s_1$ and $s_2$ yield the same potential $V_x$ in
Eq.~(\ref{e:x-eqn}). If $\eta_{in}$ is chosen such that the
contributions to the integrals from the lower boundary can be
neglected, changing $s$ from $s_1$ to $s_2$ transforms $\Tcos$
into $k^{-1}\Tsin$ and $\Tsin$ into $k\Tcos$.

\item If $s \propto \eta^\gamma$ is a pure power law and again the
contributions from the lower boundary can be neglected, $\Tcos
\propto \sqrt{k|\eta|}J_\nu(|k\eta|)$ and $\Tsin \propto
\sqrt{k|\eta|}Y_{-\nu}(|k\eta|)$, where $\nu = \gamma-1/2$ and
$J,~Y$ are Bessel functions.
\end{itemize}

We now define
\begin{equation}
 y \equiv \frac{s}{k^2} \left(x/s\right)'~.
 \label{e:def-y}
\end{equation}
Using Eq.~(\ref{e:x-eqn}), we find
\begin{equation}
 \pmatrix{y \cr y'} = -\frac{1}{k^2} \pmatrix{\frac{s'}{s} & -1 \cr
  k^2 -\left(\frac{s'}{s}\right)^2 & \frac{s'}{s}} \pmatrix{x \cr x'}~,
 \label{e:y0_and_y1}
\end{equation}
which we can invert to obtain
\begin{equation}
 \pmatrix{x \cr x'} = - \pmatrix{ \frac{s'}{s} & 1 \cr
 - k^2 +\left(\frac{s'}{s}\right)^2 & \frac{s'}{s}} \pmatrix{y \cr y'}~.
\label{e:x0_and_x1}
\end{equation}
Using the latter and Eq.~(\ref{e:x-eqn}), the evolution equation
for the variable $y$ can be derived
\begin{equation}
 y''+\left[\frac{s''}{s}-2\left(\frac{s'}{s}\right)^2\right]y+k^2 y= 0~.
 \label{e:y-eqn1}
\end{equation}
Let us now introduce
\be
 r \equiv s^{-1}\left[c_1+c_2\int^\eta
     {s}^2 d\tilde\eta\right] ~,
     \label{e:def-r}
\ee
so that
\be
 r' =  -\frac{{s}'}{s} r + c_2 {s} ~,
 \qquad
 r'' = -\left[\frac{{s}''}{{s}}
   -2\left(\frac{{s}'}{s}\right)^2\right]r~.
   \label{e:r''}
\ee
Equation~(\ref{e:y-eqn1}) then takes the simple form
\begin{equation}
 y''+ \left( k^2 - {V}_y\right) y= 0~,
 \qquad
 {V}_y \equiv \frac{r''}{r}~.
 \label{e:y-eqn2}
\end{equation}

Note that the $y$ equation obtained from a given $x$ equation
depends on our choice of $s$. Since $s_1''/s_1 = s_2''/s_2$ but
$s_1'/s_1\neq s_2'/s_2$, $V_y = -s''/s+2(s'/s)^2$ depends on this
choice. Such a ``dual variable'' $y$ can also be found if $k^2$ is
modified into $\Upsilon(\eta)k^2$, the expressions just become
somewhat more complicated. Choosing $x=v$ and $s=z_1$,
Eqs.~(\ref{e:y0_and_y1}), (\ref{e:x0_and_x1}) just reproduce the
relations (\ref{e:vu-relation}), (\ref{e:uv-relation}) where $y=u$
and $r=\theta$.  During a power law evolution of the scale factor,
we have $z_1 = \left|\eta/\eta_s\right|^q$, $z_2 =
\left|\eta/\eta_s\right|^{1-q}$, $\theta_1 =
\left|\eta/\eta_s\right|^{-q}$ and  $\theta_2 =
\left|\eta/\eta_s\right|^{1+q}.$

As we have seen in the previous section, on large scales,
$|k\eta|\ll 1$, the general solution of Eq.~(\ref{e:x-eqn}) is to
lowest order of the form
\be
 x = {\cal A}\left|\frac{\eta}{\eta_s}\right|^\gamma
   + {\cal B}\left|\frac{\eta}{\eta_s}\right|^{1-\gamma}
   + {\cal O}(|k\eta|^2)~,
\label{e:x-superHubble}
\ee
where one obtains from Eq.~(\ref{e:x-sol-neg})
\bea \label{e:def-alpha-beta}
  \alpha &=& {\cal A}
        + {\cal B}\left|\frac{\eta_{in}}{\eta_s}\right|^{1-2\gamma}~,\\
  \beta &=& -{{\cal B}(1-2\gamma)\over k\eta_s }~. \label{e:beta}
\eea

To discuss what might happen during a transition we now assume
that for a given $s=s_1$ or $s_2$ the solution
(\ref{e:sol-Tsin-Tcos}) can be continued through the transition to
the radiation era, the only effect of the transition being a
modification of $s$ which interpolates from
\be
 s = |\eta/\eta_s|^{\gamma_-} \quad\mbox{ for } \quad\eta\ll -\eta_s ~,
\ee
to
\be
 s = |\eta/\eta_s|^{\gamma_+} \quad\mbox{ for } \quad\eta\gg \eta_s ~,
\ee
without passing through zero. At some time $\eta_s\ll \eta\ll 1/k$
in the radiation era, we still can approximate the $\Tcos$ and
$\Tsin$ integrals by the first term in their series expansion
(\ref{e:def-Tcos},\ref{e:def-Tsin}). Integrating the first term in
Eq.~(\ref{e:def-Tsin}), we obtain
\bea
 x &=&\left[\alpha +{\beta k\eta_s\over 1-2\gamma_-}
      \left(\Delta_T -{2-2\gamma_- -2\gamma_+\over1-2\gamma_+}
      +\left|\frac{\eta_{in}}{\eta_s}\right|^{1-2\gamma_-} \right)\right]
      \left({\eta\over\eta_s}\right)^{\gamma_+} \nonumber \\
  && + {\beta k\eta_s\over
   1-2\gamma_+}\left({\eta\over \eta_s}\right)^{1-\gamma_+}  ~ .
 \label{e:x-radiation-1}
\eea
Here
\be
 \Delta_T = {1-2\gamma_-\over \eta_s}\int_{-\eta_s}^{\eta_s} s^{-2}d\tilde\eta
\ee
comes from the contribution to the integral during the transition.
We always assume that the free normalization of $s$ is chosen such
that $s(-\eta_s) =1$. The dimensionless, $k$-independent constant
$\Delta_T$ is then the only quantity that incorporates our
ignorance of the true form of the transition. Although its typical
order of magnitude is ${\cal O}(\Delta_T) = {\cal O}(1)$, we will
mention explicit limits on $\Delta_T$ as we go along. With the
above expressions for $\alpha$ and $\beta$ we have
\be \label{e:x-radiation-2}
 x =\left[{\cal A} - {\cal B}\left(\Delta_T -
   {2-2\gamma_--2\gamma_+\over 1-2\gamma_+}\right)\right]
 \left({\eta\over\eta_s}\right)^{\gamma_+} - {\cal B}{1-2\gamma_-\over
   1-2\gamma_+}\left({\eta\over \eta_s}\right)^{1-\gamma_+}~,
\ee
which is independent of $\eta_{in}$ as it should be.

In what follows we will study eight different cases and compute
the resulting spectra. For the first four cases we shall assume
that $u$ remains regular throughout. Recalling the notation that
$\theta_{-}$ refers to the collapsing phase and $\theta_{+}$ to
the expanding phase in Eqs.~(\ref{e:u-eqn}), (\ref{e:def-theta}),
we shall consider the following possibilities: $\theta_{1-}
=|\eta/\eta_s|^{-q}$ goes over smoothly into $\theta_{1+} =
(\eta/\eta_s)^{-1}$ (case 1); $\theta_{1-}$ goes over smoothly
into $\theta_{2+} =(\eta/\eta_s)^2$ (case 2);
$\theta_{2-}=|\eta/\eta_s|^{1+q}$ goes over smoothly into
$\theta_{1+}$ (case 3) and $\theta_{2-}$ goes over smoothly into
$\theta_{2+}$ (case 4). We shall then study the equivalent cases
for $v$ with $\theta$ replaced by $z$ in Eqs.~(\ref{e:v-eqn}),
(\ref{e:def-z}). We are mainly interested in a contracting
pre-big-bang phase, $q>0$, but the results derived here are valid
also for $-1/2<q$.

\medskip
\noindent
{\bf Case 1:}\hspace{2mm} $\gamma_- = -q$,\quad $\gamma_+ =-1$.\\
Here we have ${\cal A} = A_-$ and ${\cal B} = B_-$, hence
\bea
 u &\simeq& k^{-3/2}\left\{\left[(k\eta_s)^{-q}
   -(k\eta_s)^{1+q}\left(\Delta_T-{2(2+q)\over 3}\right)\right]
   \left({\eta\over\eta_s}\right)^{-1}  \right. \nonumber \\
   && \quad\quad\quad  \left.~ -{1+2q\over 3}(k\eta_s)^{1+q}
    \left({\eta\over\eta_s}\right)^2 \right\} ~.
\eea
If $\Delta_T$ is of order unity, or more precisely if
\be \label{e:condition-1}
 \Delta_T ~ \lsim ~ (k\eta_s)^{-(1+2q)}~,
\ee
 the resulting spectrum as well as the amplitude
does not depend on $\Delta_T$ and we have
\be
 |u|^2k^3 \simeq \left\{ \begin{array}{lcl}
  (k\eta_s)^{-2q}\left({\eta\over \eta_s}\right)^{-2} & \mbox{ for }
     k<k_u(\eta) ~& n =1-2q~, \\
 \left({1+2q\over 3}\right)^2(k\eta_s)^{2+2q}\left({\eta\over
     \eta_s}\right)^4 & \mbox{ for }
     k> k_u(\eta) ~& n =3+2q ~,  \end{array} \right.
\ee
where
\be
 k_u(\eta) \simeq \eta^{-1}
 \left({\eta\over\eta_s}\right)^{(2q-2)/(2q+1)}
\ee
is the wave number where we see a kink in the spectrum. For a
value of $q$ in the regime of our primary interest, $-1/2<q<1$,
the exponent $(2q-2)/(2q+1)$ is negative and $k_u(\eta)\ll
\eta^{-1}$ especially at late times, $\eta\gg\eta_s$. Hence, in
this case the spectral index relevant for the observed
anisotropies in the cosmic microwave background (CMB) is $n=3+2q$,
a steep blue spectrum. In reaching this conclusion, we have used
the fact that in the radiation era,
\bea
 \Psi &\simeq& {\sqrt{2}\eta_s\over M_P\eta^2}u~, \\
 |\Psi|^2k^3 &\simeq&  \left({M_s \over M_P}\right)^{2}
 (k\eta_s)^{2+2q}~\mbox{ for }
     k> k_u(\eta)~, \quad n=3+2q ~,
\eea
where we have introduced the transition mass scale,
$M_s=\eta_s^{-1}$. For transitions from contraction to expansion,
$q>0$, the amplitude of these fluctuations is therefore far too
low to be of any relevance for cosmologically interesting scales,
$k\simeq \eta_0^{-1}$. (Furthermore, the spectral index $n=3+2q$
is not consistent with observations.)

\medskip
\noindent
{\bf Case 2:}\hspace{2mm} $\gamma_- = -q$,\quad $\gamma_+ =2$.\\
Since $\gamma_-$ is the same as in case~1, $\cal A$ and $\cal B$
remain unchanged. From Eq.~(\ref{e:x-radiation-2}), we find
\bea
u &\simeq& k^{-3/2}\left\{\left[(k\eta_s)^{-q} -
    (k\eta_s)^{1+q}\left(\Delta_T-{2(1-q)\over3}\right)\right]
    \left({\eta\over\eta_s}\right)^{2}  \right.
\nonumber \\  && \quad\quad\quad  \left. ~+{1+2q\over
    3}(k\eta_s)^{1+q}\left({\eta\over\eta_s}\right)^{-1} \right\} ~,
\eea
so that
\bea
|u|^2k^3 &\simeq&   (k\eta_s)^{-2q}\left({\eta\over \eta_s}\right)^4,\\
 |\Psi|^2k^3 &\simeq&  \left({M_s\over M_P}\right)^2(k\eta_s)^{-2q}~,
 \quad n=1-2q~.
\eea
For this result to apply, the condition on $\Delta_T$,
Eq.~(\ref{e:condition-1}) must be satisfied. If this case is
realized and if $0<q\ll 1$, a scale invariant spectrum will be
obtained. Its amplitude is determined by the transition scale
which should be about 5 orders of magnitude below the Planck
scale.

\medskip
\noindent
{\bf Case 3:}\hspace{2mm} $\gamma_- = 1+q$,\quad $\gamma_+ =-1$.\\
According to Eqs.~(\ref{e:x-superHubble}) and
(\ref{e:u-superHubble}), we now have
 ${\cal A} = B_-$ and ${\cal B} = A_-$.
This leads to
\bea
u &\simeq& k^{-3/2}\left\{\left[(k\eta_s)^{1+q} -
    (k\eta_s)^{-q}\left(\Delta_T-{2(1-q)\over3}\right)\right]
    \left({\eta\over\eta_s}\right)^{-1}  \right.
  \nonumber \\  && \quad\quad\quad  \left. ~+{1+2q\over3}
  (k\eta_s)^{-q}\left({\eta\over\eta_s}\right)^{2} \right\}~.
\eea
We thus obtain
\bea
 |u|^2k^3 & \simeq &   (k\eta_s)^{-2q}
 \left({\eta\over \eta_s}\right)^4~,\\
 |\Psi|^2k^3 & \simeq &  \left({M_s\over M_P}\right)^2
 (k\eta_s)^{-2q}~, \quad n=1-2q~,
\eea
as in case~2. Here this spectrum is obtained without any condition
on $\Delta_T$ having to be satisfied, although for the correct
amplitude to be obtained, we need $\Delta_T \lsim
(\eta/\eta_s)^3/3$.

\medskip
\noindent
{\bf Case 4:}\hspace{2mm} $\gamma_- = 1+q$,\quad $\gamma_+ =2$. \\
Again, we have we have ${\cal A} = B_-$ and ${\cal B} = A_-$ and
so we obtain
\bea
u &\simeq& k^{-3/2}\left\{\left[(k\eta_s)^{1+q} -
    (k\eta_s)^{-q}\left(\Delta_T-{2(2+q)\over3}\right)\right]
    \left({\eta\over\eta_s}\right)^{2}  \right.
  \nonumber \\  && \quad\quad\quad  \left. ~-{1+2q\over3}
  (k\eta_s)^{-q}\left({\eta\over\eta_s}\right)^{-1} \right\}~,
\eea
with
\bea
|u|^2k^3 &\simeq&
(k\eta_s)^{-2q}\left(\Delta_T-{2(2+q)\over3}\right)^2
  \left({\eta\over \eta_s}\right)^4~,\\
|\Psi|^2k^3 &\simeq&  \left({M_s\over M_P}\right)^2(k\eta_s)^{-2q}
\left(\Delta_T-{2(2+q)\over3}\right)^2~, \quad n=1-2q ~.
\eea
Again, we obtain a scale invariant spectrum if $q\ll 1$, but in
this case the amplitude depends on the details of the transition
given by $\Delta_T$.

We now repeat this analysis considering the alternative variable
$v$ with 'pump field' $s=z_1$ or $z_2$.

\medskip
\noindent
{\bf Case 1:}\hspace{2mm} $\gamma_- = q$,\quad $\gamma_+ =1$.\\
According to Eqs.~(\ref{e:x-superHubble}) to (\ref{e:beta}) we
have  ${\cal A} = D_-$ and ${\cal B} = C_-$, leading to
\bea
 v &=& \sqrt{\eta_s}\left\{\left[(k\eta_s)^{q-1/2}
     -(k\eta_s)^{1/2-q}\left(\Delta_T-2q\right)\right]\left({\eta\over
       \eta_s}\right)  +(1-2q)(k\eta_s)^{1/2-q}\right\}~,  \\
|v|^2k^3 &\simeq& \left\{\begin{array}{ll}
 \eta_s^{-2}(k\eta_s)^{2+2q}\left({\eta\over \eta_s}\right)^2
 &  \mbox{ for } q\leqslant1/2~,\\
   \eta_s^{-2}(k\eta_s)^{4-2q}
   \left(\Delta_T-2q\right)^2\left({\eta\over \eta_s}\right)^2
 &  \mbox{ for } q> 1/2~, \end{array} \right. \\
|\zeta|^2k^3 &\simeq & \left\{\begin{array}{ll}
  \left({M_s\over M_P}\right)^2 (k\eta_s)^{2+2q},
 & n=3+2q  ~\mbox{ if } ~q\leqslant1/2~,\\
 \left({M_s\over M_P}\right)^2 (k\eta_s)^{4-2q}\left(\Delta_T-2q\right)^2,
 & n=5-2q  ~\mbox{ if }~ q > 1/2~. \end{array} \right.
\eea
If $q<1/2$ we must require $\Delta_T \lsim (k\eta_s)^{2q-1}$ for
our result to apply. Note also that the amplitude of the spectrum
depends on the details of the transition given by $\Delta_T$ when
$q>1/2$.

\medskip
\noindent
{\bf Case 2:}\hspace{2mm} $\gamma_- = q$,\quad $\gamma_+ =0$.\\
Again, we have  ${\cal A} = D_-$ and ${\cal B} = C_-$, which
yields
\bea
 v &=& \sqrt{\eta_s}\left\{\left[(k\eta_s)^{q-1/2}
     -(k\eta_s)^{1/2-q}\left(\Delta_T-2+2q\right)\right]
     -(1-2q)(k\eta_s)^{1/2-q}\left({\eta\over
       \eta_s}\right)\right\}~,  \\
 |v|^2k^3 &\simeq & \left\{\begin{array}{ll}
   \eta_s^{-2}(k\eta_s)^{4-2q}\left({\eta\over \eta_s}\right)^2 &
    \mbox{ if }~ q > 1/2 ~\mbox{ or }~ k >  k_{v_1}(\eta)~,\\
 \eta_s^{-2}(k\eta_s)^{2+2q}  & \mbox{ if }~
 q \leqslant 1/2 ~\mbox{ and }~ k < k_{v_1}(\eta)~,
 \end{array}\right. \\
 |\zeta|^2k^3 &\simeq & \left\{\begin{array}{ll} \left({M_s\over
 M_P}\right)^2(k\eta_s)^{4-2q}~,
  & n=5-2q  ~\mbox{ if }~ q > 1/2 ~\mbox{ or }~ k>k_{v_1}~, \\
 \left({M_s\over M_P}\right)^2(k\eta_s)^{2+2q}\left({\eta\over
 \eta_s}\right)^{-2}~,
 & n=3+2q  ~\mbox{ if }~ q \leqslant 1/2 ~\mbox{ and }~ k <  k_{v_1}~.
       \end{array}\right.
\eea
For the result to apply when $q>1/2$ it requires $\Delta_T<
\eta/\eta_s$. As in Case 1 of the $u$ field, there is a kink in
the spectrum with the wave number of the kink for the case
$q\leqslant1/2$ being
\be
k_{v_1}(\eta) =
\eta^{-1}\left({\eta\over\eta_s}\right)^{-2q/(1-2q)}~,
\ee
which is always smaller than the Hubble radius, $\eta^{-1}$, for
the relevant values of $q$, $0<q<1/2$. Only for very small values
of $q$, this kink in the spectrum lies very close to the Hubble
radius and is not visible.

\medskip
\noindent
{\bf Case 3:}\hspace{2mm} $\gamma_- = 1-q$,\quad $\gamma_+ =1$.\\
Here we have  ${\cal A} = C_-$ and ${\cal B} = D_-$, hence
\bea
 v &=& \sqrt{\eta_s}\left\{\left((k\eta_s)^{1/2-q}
     -(k\eta_s)^{q-1/2}[\Delta_T -2+2q]\right)\left({\eta\over
       \eta_s}\right) - (k\eta_s)^{q-1/2}(1-2q)\right\}~,  \\
|v|^2k^3 &\simeq&\left\{ \begin{array}{ll}
  \eta_s^{-2}(k\eta_s)^{4-2q}\left({\eta\over
       \eta_s}\right)^2 &\mbox{ if }~ q> 1/2~, \\
 \eta_s^{-2}(k\eta_s)^{2+2q}\left({\eta\over
       \eta_s}\right)^2[\Delta_T -2+2q]^2 & \mbox{ if }~
       q\leqslant1/2~,
 \end{array} \right. \\
 |\zeta|^2k^3 &\simeq & \left\{ \begin{array}{ll} \left({M_s\over
 M_P}\right)^2 (k\eta_s)^{4-2q}~,
     & n=5-2q ~\mbox{ if }~ q> 1/2~, \\
 \left({M_s\over M_P}\right)^2 (k\eta_s)^{2+2q}[\Delta_T -2+2q]^2~,
 & n=3+2q  ~\mbox{ if }~ q\leqslant1/2~.
 \end{array} \right.
\eea
For $q>1/2$ the amplitude of the resulting fluctuations does not
depend on the details of the transition while it does depend on it
for $q\leqslant1/2$.

\medskip
\noindent
{\bf Case 4:}\hspace{2mm} $\gamma_- = 1-q$,\quad $\gamma_+ =0$.\\
Here again we have  ${\cal A} = C_-$ and ${\cal B} = D_-$, so that
\bea
 v &=& \sqrt{\eta_s}\left\{(k\eta_s)^{1/2-q}
     -(k\eta_s)^{q-1/2}[\Delta_T-2q] + (1-2q)(k\eta_s)^{q-1/2}
   \left({\eta\over \eta_s}\right)\right\} ~, \\
 |v|^2k^3 &\simeq&\left\{ \begin{array}{ll}
  \eta_s^{-2}(k\eta_s)^{4-2q} &\mbox{ if }~q> 1/2~
  \mbox{ and }~k<k_{v_2}~, \\
 \eta_s^{-2}(k\eta_s)^{2+2q}\left({\eta\over
       \eta_s}\right)^2 & \mbox{ if }~q\leqslant 1/2~
       \mbox{ or }~k<k_{v_2}~,
 \end{array} \right. \\
 |\zeta|^2k^3 &\simeq & \left\{ \begin{array}{ll} \left({M_s\over
 M_P}\right)^2 (k\eta_s)^{4-2q}
 \left({\eta\over\eta_s}\right)^{-2},
 & n= 5-2q ~\mbox{ if }~ q> 1/2  \mbox{ and } k<k_{v_2}~, \\
 \left({M_s\over M_P}\right)^2 (k\eta_s)^{2+2q}~,
 & n= 3+2q ~ \mbox{ if }~ q\leqslant  1/2
             \mbox{ or } k<k_{v_2}~,
 \end{array} \right.
\eea
where here
\be
 k_{v_2}(\eta) \simeq\eta^{-1}\left({\eta\over\eta_s}
 \right)^{(2-2q)/(1-2q)} ~.
\ee
For $1/2<q<1$ and $k_{v_2} <\eta^{-1}$, a kink from $n=3+2q$ to
the final spectrum $n=5-2q$ is present in the spectral
distribution.

\bigskip
In Tables \ref{t:n-u}--\ref{t:n-v>1/2} we summarize the results of
our analysis:

\begin{table}[ht]
 \begin{center} {\bf The spectral index for a transition
     with regular $u$ and  $q>-1/2$}  \vspace{1mm}\\
  \begin{tabular}{|c|c|c|c|c|c|c|}
    \hline
    case & $\gamma_{-}$ & $\gamma_{+}$ & kink? &
    stable? & ampl. depends & $n$\\
    &&&&& on transition? & \\
    \hline\hline
    1 & $-q$  & $ -1 $ & yes & no  & no & $3+2q$ \\
    2 & $-q$  & $ 2  $ &  no & yes & no & $1-2q$\\
    3 & $1+q$ & $ -1 $ &  no & yes & no & $1-2q$\\
    4 & $1+q$ & $ 2 $ &  no & yes & yes& $1-2q$\\
    \hline
  \end{tabular}
 \end{center}
\caption[Post-big bang spectral indices] {\label{t:n-u}Here we
summarize the post-big-bang spectral indices as a function of the
pre- and post-big-bang exponent of the pump field, if $u$ is
regular through the transition.  The mild requirement on the
transition ($\Delta_T$) for this to hold is mentioned in the
text.}
\end{table}

\begin{table}[ht]
 \begin{center} {\bf The spectral index for a transition
     with regular $v$ and  $q\leqslant 1/2$} \vspace{1mm}\\
  \begin{tabular}{|c|c|c|c|c|c|c|}
    \hline
    case & $\gamma_{-}$ & $\gamma_{+}$ & kink? &
    stable? & ampl. depends & $n$\\
    &&&&& on transition? & \\
    \hline\hline
    1 & $ q$ & $ 1 $  &  no & yes & no & $3+2q$\\
    2 & $ q$ & $ 0  $ & yes & no & no & $5-2q$ \\
    3 & $1-q$ & $ 1  $ &  no & yes & yes & $3+2q$\\
    4 & $1-q$ & $ 0  $ &  no & yes & no & $3+2q$\\
    \hline
  \end{tabular}
 \end{center}
\caption[Post-big bang spectral indices] {\label{t:n-v<1/2}Here we
summarizes the post-big-bang spectral indices as a function of the
pre- and post-big-bang exponent of the pump field, if $v$ is
regular through the big bang and $q\leqslant 1/2$. The logarithmic
corrections at $q=1/2$ are neglected. The mild requirement on the
transition ($\Delta_T$) for this to hold is mentioned in the
text.}
\end{table}

\begin{table}[ht]
 \begin{center} {\bf The spectral index for a transition
     with regular $v$ and  $q> 1/2$}  \vspace{1mm} \\
  \begin{tabular}{|c|c|c|c|c|c|c|}
    \hline
    case & $\gamma_{-}$ & $\gamma_{+}$ & kink? &
    stable? & ampl. depends & $n$\\
    &&&&& on transition? & \\
    \hline\hline
    1 & $ q$  & $ 1 $ &  no & yes & yes & $5-2q$\\
    2 & $ q$  & $ 0 $ & no  & yes & no  & $5-2q$ \\
    3 & $1-q$ & $ 1 $ &  no & yes & no  & $5-2q$\\
    4 & $1-q$ & $ 0 $ & yes & no  & no  & $3+2q$\\
    \hline
  \end{tabular}
 \end{center}
\caption[Post-big bang spectral indices] {\label{t:n-v>1/2}Here we
summarizes the post-big-bang spectral indices as a function of the
pre- and post-big-bang exponent of the pump field, if $v$ is
regular through the big bang and $q>1/2$. The mild requirement on
the transition ($\Delta_T$) for this to hold is mentioned in the
text.}
\end{table}

%\clearpage

From Eqs.~(\ref{e:def-u-Psi}) and (\ref{e:Psi-eqn-reduced}) it is
clear that during the radiation dominated era, inside the Hubble
radius, $k\eta\gg 1$, the Bardeen potential oscillates and its
amplitude decays like $1/\eta^2$, whereas during the matter
dominated era, the Bardeen potential remains constant also inside
the Hubble radius. Therefore, a change in the spectral index close
to the Hubble radius crossing is not visible for scales which
cross the Hubble radius in the radiation dominated era. This
remark concerns mainly the kink in case~2 of
Table~\ref{t:n-v<1/2}.

A kink in the spectral distribution arises only in the unlikely
situation where the growing mode of the pre-big-bang phase is
fully converted into the decaying mode after the transition, and
one has to wait a sufficiently long time $[\sim \eta(k)]$ for the
decaying mode to decay and the final growing mode to dominate. It
is only in such a situation that the final spectral index does not
correspond to the naive expectation from the pre-big-bang phase.
Finally, we note that a kink is always associated with an
instability of the spectrum. The issue of stability will be
discussed in Sec.~\ref{s:toy} where we model the regular behavior
of $u$ and $v$ through simple toy models. There we shall see that
a slight modification in the transition can change the spectral
index $n=3+2q$ into $n=1-2q$ if $u$ passes through the transition
regularly and $5-2q$ into $3+2q$ if $v$ is regular and $q\leqslant
1/2$, correspondingly $3+2q$ into $5-2q$ if $v$ is regular and $q
> 1/2$.

This brings us to one of the key results of this paper, a
prediction of the spectral index arising from different conditions
on $q$ and the regularity of the $u$ and $v$ fields:
\be \label{e:spectral-indices}
 n = \left\{ \begin{array}{lcl}
  1-2q & \mbox{ if } & q>-1/2
        ~\mbox{ and $u$ is regular and stable},\\
  3+2q & \mbox{ if } & q\leqslant 1/2
        ~\mbox{ and $v$ is regular and stable}, \\
  5-2q & \mbox{ if } & q>1/2
        ~\mbox{ and $v$ is regular and stable}, \\
  3+2q & \mbox{ if } & q\leqslant -1/2.
\end{array} \right.
\ee
We have not treated explicitly the simple case $q< -1/2$ above,
but this can be done exactly along the same lines as the other
cases. From Eq.~(\ref{e:spectral-indices}), we see that a
scale-invariant spectrum is obtained for $q\simeq 1$ (standard
inflation), or if $u$ is regular and $0 < q \ll 1$, or if $v$ is
regular and $q=2$. For this latter case, however, we shall see in
Sec.~\ref{ss:amplitude} that perturbations grow large during the
contracting phase and therefore linear perturbation theory breaks
down. Furthermore, such a collapsing universe with $q=2$ has been
shown to be unstable~\cite{Heard:2002dr}.

Clearly, in a transition from contraction to expansion, it cannot
be that both $u$ and $v$ are regular and stable if $q>-1/2$. Only
in an inflationary transition with $q\leqslant-1/2$ do we find
$n=3+2q$ for both $u$ and $v$. In this case it is expected that
both variables transit in a regular stable fashion from inflation
to the radiation era. The resulting spectral index does not depend
on the variable with which the calculation is performed. The
situation for arbitrary values of $q$ is shown in
Fig.~\ref{f:indices_pbb}.

\begin{figure}[t]
 \begin{center}
 \includegraphics[width=0.45\linewidth]{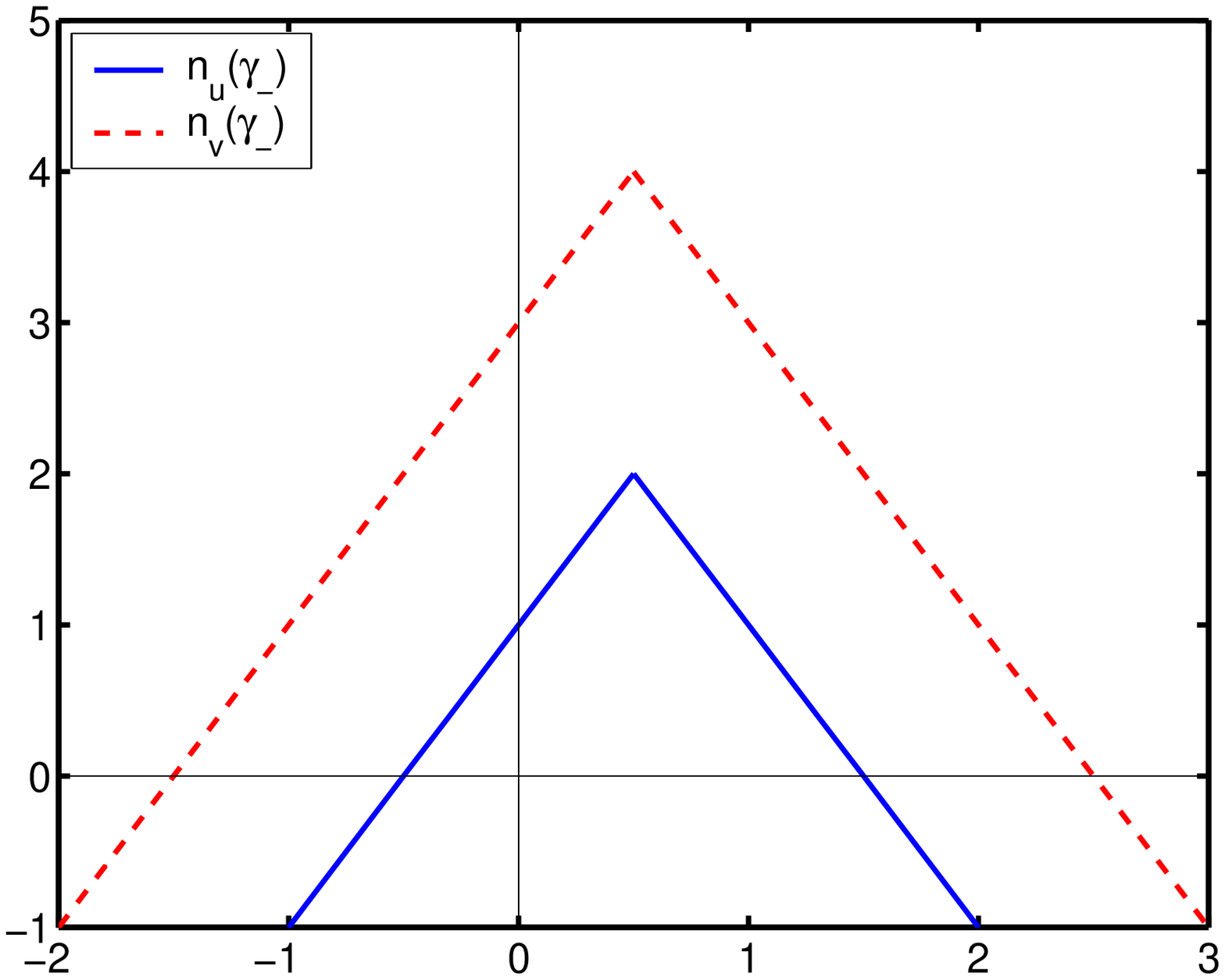}
 \quad
 \includegraphics[width=0.45\linewidth]{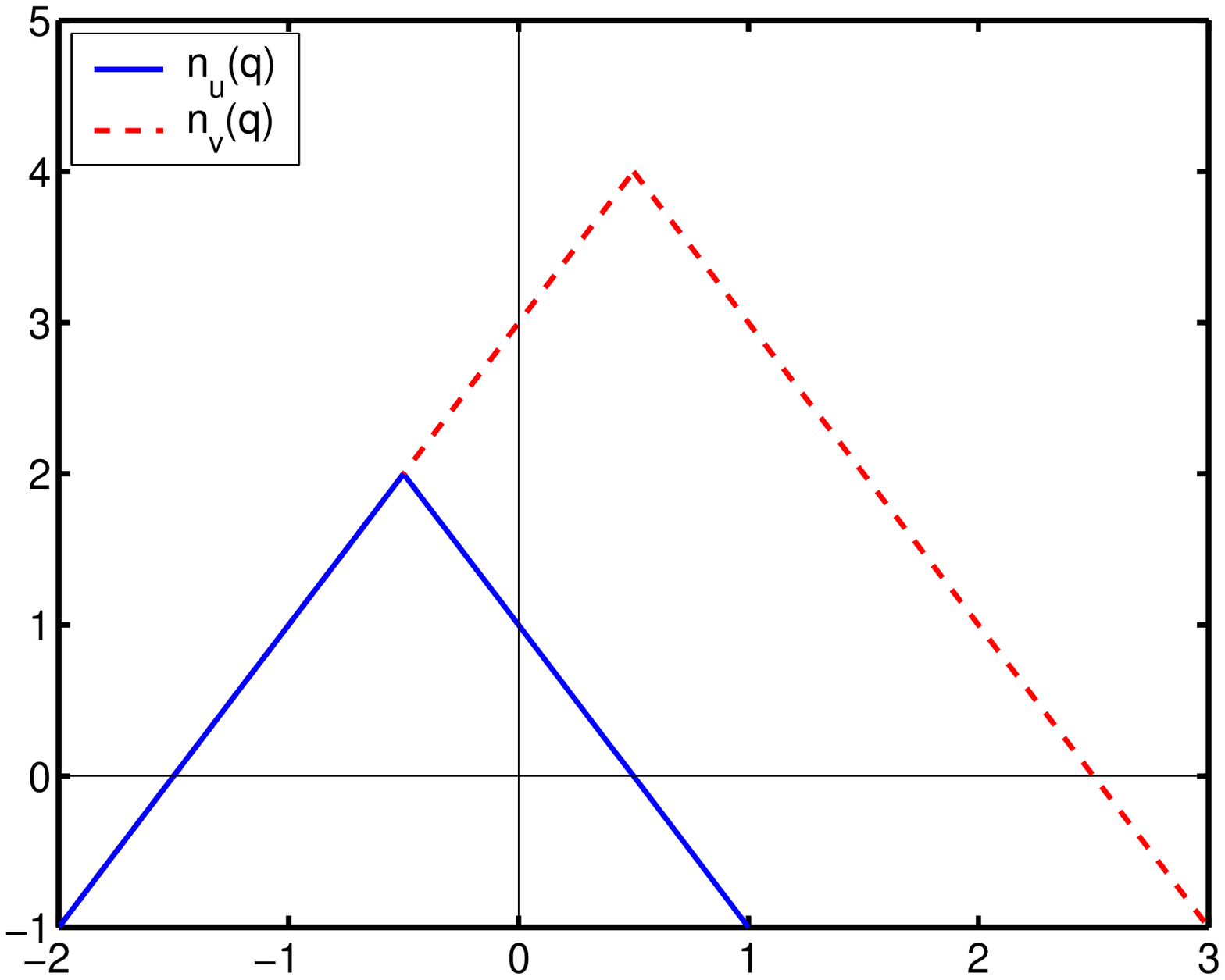}
 \caption
 {\label{f:indices_pbb}
Here we illustrate the dominant spectral indices $n_u$ (solid) and
$n_v$ (dashed) on super-Hubble scales which are obtained after a
``stable'', regular transition of the corresponding variable to
the radiation era at sufficiently late time. This is also the
resulting spectral index for the Bardeen potential on
cosmologically relevant scales. The left panel shows the indices
as a function of $\gamma_{-}$, the exponent of the pump field
during the pre-big-bang phase. The right panel shows them as a
function of $q$, the exponent of the contraction (expansion) law
before the big bang. We have used $\gamma_{u_{-}}=-q$ and
$\gamma_{v_{-}}=q$.}
 \end{center}
\end{figure}

\section{Fast toy model transitions}
\label{s:toy}

In earlier
work~\cite{Cartier:1999vk,Cartier:2001is,Tsujikawa:2002qc} a
transition from contraction to expansion was achieved via a
combination of first order corrections in the string scale
$\alpha'$ and/or the string coupling $g_s$.
In~\cite{Cartier:2001is} a modified perturbation equation for $v$
was derived using this framework, and a spectral index $n=3+2q$
was obtained. This yields $n=4$ for the case dilaton-driven string
cosmology~\cite{Cartier:2001is} where $q=1/2$, and $n=3$ for the
ekpyrotic model~\cite{Tsujikawa:2002qc} where $q\sim 0$. However,
although calculations have been performed with $v$, it remains to
be shown that $u$ cannot pass through the transition regularly (to
first order). Unfortunately, even though the perturbation
equations of~\cite{Cartier:2001is} are very complicated, they are
probably not realistic. It is clear that at a time where first
order corrections become important, higher order corrections are
likely to be relevant and the real behavior of the perturbations
might differ significantly from the results obtained in the work
cited above. In this sense, pre-big-bang models including first
order corrections are only toy models.

In this section we confirm our generic findings concerning the
spectral index associated with the relevant $u$ and $v$ fields by
numerically solving a simple toy model. We do not insist on a
physically well motivated transition. Rather we artificially
define a regular scale factor so that it agrees with a contracting
Friedmann universe with contraction exponent $q$ at $\eta\ll
-\eta_s$ and with a radiation dominated universe at $\eta\gg
\eta_s$. In the region in between, the scale factor smoothly
evolves between contraction and expansion.

\subsection{The background}
For the exact form of the regularized background scale factor we
choose
\bea
 a(\eta) &=& \left[\left(\eta/\eta_s\right)^2
 +\epsilon\right]^{\tilde{q}/2}~,\\
 \tilde{q}(\eta) &=& q\iota +(1-\iota) ~,\\
 \iota(\eta) &=& 1/2 -{1\over \pi}\arctg(\eta/\eta_s) ~,\label{e:def-iota}
\eea
where $0 < \epsilon\ll 1$. The function $\iota(\eta)$ could be
replaced by any $C^\infty$ function which quickly interpolates
between $1$ for $\eta\ll -\eta_s$ to $0$ for $\eta\gg \eta_s$.
Clearly, this universe contracts like $|\eta|^q$ for $\eta\ll
-\eta_s$ and expands like a radiation dominated universe $a\propto
\eta$ for $\eta\gg \eta_s$. Furthermore, $a$, ${\cal H}$, ${\cal
H}'$ and ${\cal H}^2-{\cal H}'$ are all regular, even analytic in
the vicinity of the transition, $\eta=0$. The behavior of the
relevant background quantities for our model are shown in
Fig.~\ref{f:short-transition}.

\begin{figure}[t]
 \begin{center}
 \includegraphics[width=0.8\linewidth]{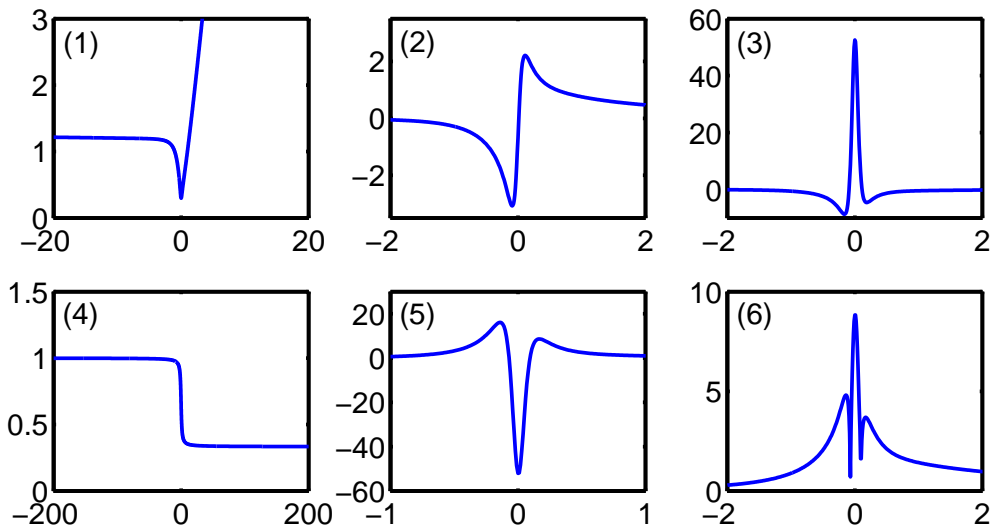}
 \caption[Toy model for a short transition]
 {\label{f:short-transition}
This figure illustrates the evolution of various background
quantities scaled in such a way that they are dimensionless (i.e.,
invariant under a change in $\eta_s$). As functions of
$\eta/\eta_s$, they are (1) the scale factor $a(\eta)$, (2) the
Hubble rate ${\cal H}\eta_s$ and (3) its first time derivative
${\cal H}'\eta_s^2$. On the bottom line, they are (4) the
pre-factor of the comoving wave number $\Upsilon$, (5) the
potential of the perturbation variable $u$, i.e.,
$V_u\eta_s^2=\left({\cal H}^2-{\cal H}'\right)\eta_s^2$ and (6)
its rescaled square root $\sqrt{|V_u|/\Upsilon}~\eta_s$. The
parameters used for these figures are $\epsilon = 10^{-2}$ and $q
= 5 \cdot 10^{-2}$.}
 \end{center}
\end{figure}

\subsection{A regular transition in the perturbation variable $u$}

We first consider a regular transition for $u$. In the regime
where the scale factor is a simple power law, $|\eta|\gg \eta_s$,
the $u$ equation is given by Eq.~(\ref{e:u-eqn-reduced}). During
this regime, the $u$ potential is simply $V_u = q(q+1)/\eta^2$. In
order to regularize this potential during the transition, we
impose $V_u =\theta''/\theta$ throughout, where
\bea
 \theta(\eta) &=& \left[ \left(\eta/\eta_s\right)^2
 +\epsilon\right]^{\tilde{\gamma}/2}~,\\
 \tilde{\gamma}(\eta) &=& \iota\gamma_- +(1-\iota)\gamma_+ ~.
 \label{e:def-gamma}
\eea
This ensures us that the pump field $\theta(\eta)$ remains regular
during the whole evolution and reduces to the power law asymptotic
regimes, $\theta \to |\eta/\eta_s|^{\gamma_-}$ for
$\eta\ll-\eta_s$ and $\theta \to (\eta/\eta_s)^{\gamma_+}$ for
$\eta\gg\eta_s$.

The prefactor of the comoving wave number is $\Upsilon=1$ during
the scalar field dominated pre-big-bang phase and
$\Upsilon=c_s^2=1/3$ in the radiation dominated era. We regularize
$\Upsilon$ during the transition via
\be \label{e:def-Upsilon}
   \Upsilon(\eta) = {1\over 3}\left(2\iota + 1\right) ~.
\ee
It is worth mentioning at this point that the results we have
obtained appear to be quite insensitive to how $\Upsilon(\eta)$ is
modelled. To choose the pump field of case~1 of
Sec.~\ref{ss:perturbation-u-v}, $\theta=\theta_1$, requires
$\gamma_-=-q$ and $\gamma_+=-1$. Similarly, setting
$\theta=\theta_2$ corresponds to case~4, with $\gamma_-=1+q$ and
$\gamma_+=2$. To obtain also the cases~2 and 3 we need $\theta$ to
interpolate from $\theta_1$ to $\theta_2$ (case~2) and from
$\theta_2$ to $\theta_1$ (case~3), respectively. We can achieve
these behaviors using our fast interpolating function
$\iota(\eta)$ given in Eq.~(\ref{e:def-iota}). The four cases are
then obtained by the following choices:
\be \label{e:regpumpu}
 \tilde{\gamma}(\eta) = \left\{ \begin{array}{rccrcr}
  \iota(1-q)-1 & \mbox{ (case 1: } & \theta_1 \to \theta_1~,~ &  -q & \to & -1)~, \\
 -\iota(2+q)+2 & \mbox{ (case 2: } & \theta_1 \to \theta_2~,~ &  -q & \to &  2)~, \\
  \iota(2+q)-1 & \mbox{ (case 3: } & \theta_2 \to \theta_1~,~ & 1+q & \to & -1)~, \\
  \iota(q-1)+2 & \mbox{ (case 4: } & \theta_2 \to \theta_2~,~ & 1+q & \to &  2)~.
\end{array} \right.
\ee
It is easy to verify that the given functional forms have the
correct asymptotic behavior. Furthermore, they are clearly regular
throughout. The numerical results for the $u$-spectra are shown in
Figs.~\ref{f:case-1} to \ref{f:case-4}. The wave number $k$ is
given in units of the maximum amplified wave number defined by \be
\label{e:kmax}
 k_m = \max(\sqrt{|V_u|/\Upsilon})~,
\ee
which is of the order of $\eta_s^{-1}$. More precisely we have
\be
  k_m \simeq \left\{\begin{array}{r}
   8.9~\eta_s^{-1}~\mbox{ in case 1~,} \\
  12.2~\eta_s^{-1}~\mbox{ in case 2~,} \\
   3.7~\eta_s^{-1}~\mbox{ in case 3~,} \\
  15.1~\eta_s^{-1}~\mbox{ in case 4~.} \end{array}\right.
\ee

\begin{figure}[ht]
 \begin{center}
 \includegraphics[width=0.45\linewidth]{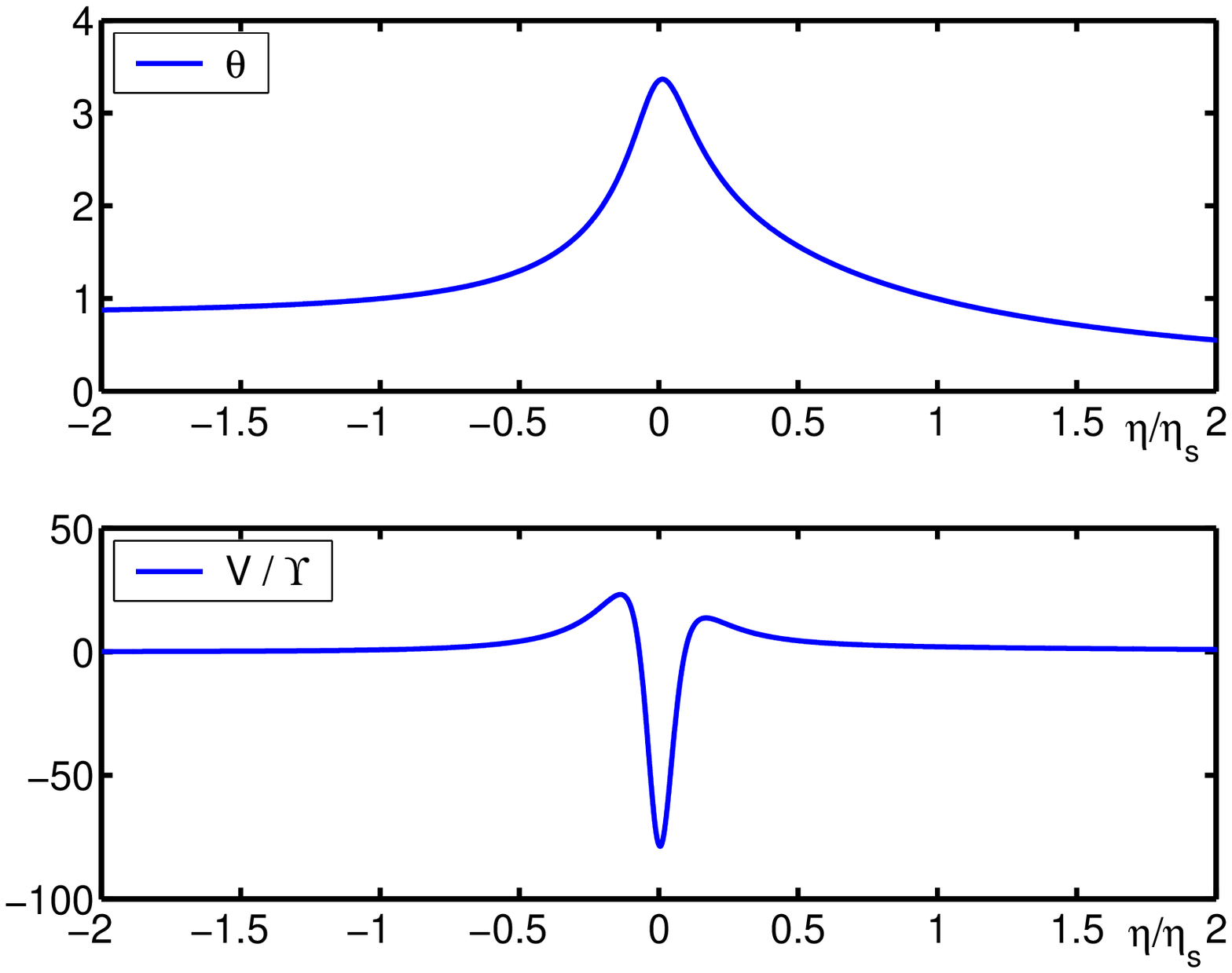}
 \quad
 \includegraphics[width=0.45\linewidth]{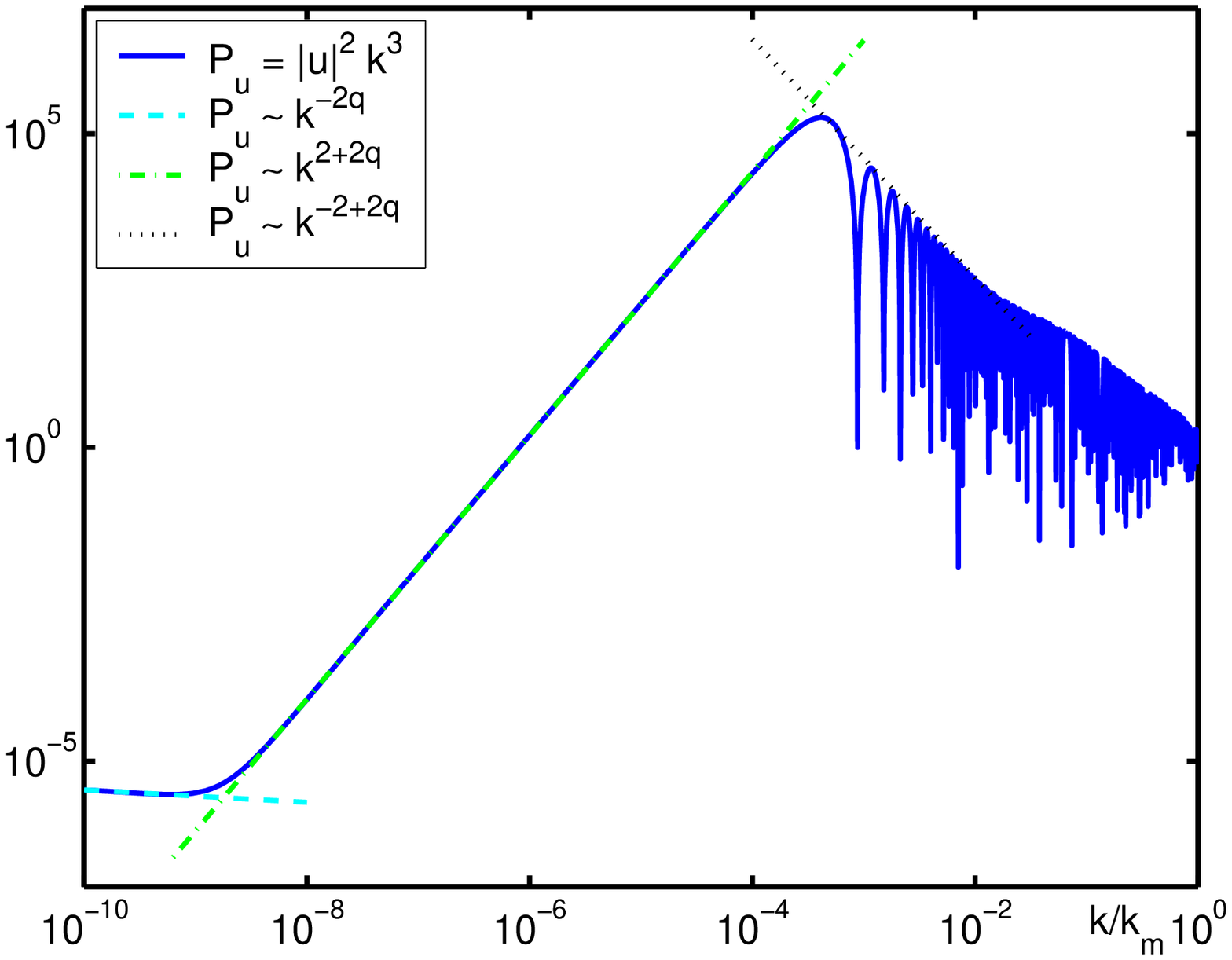}
 \caption
{\label{f:case-1} \textbf{Case 1}, $\theta_1 \to \theta_1$, using
$\tilde{\gamma}(\eta) = \iota(1-q)-1$. The $u$ potential (left)
and spectrum, $P_u=|u|^2k^3$ (right) are shown for $q=5\cdot
10^{-2}$ and
 $\epsilon=10^{-2}$.}
 \end{center}
\end{figure}

\begin{figure}[ht]
 \begin{center}
 \includegraphics[width=0.45\linewidth]{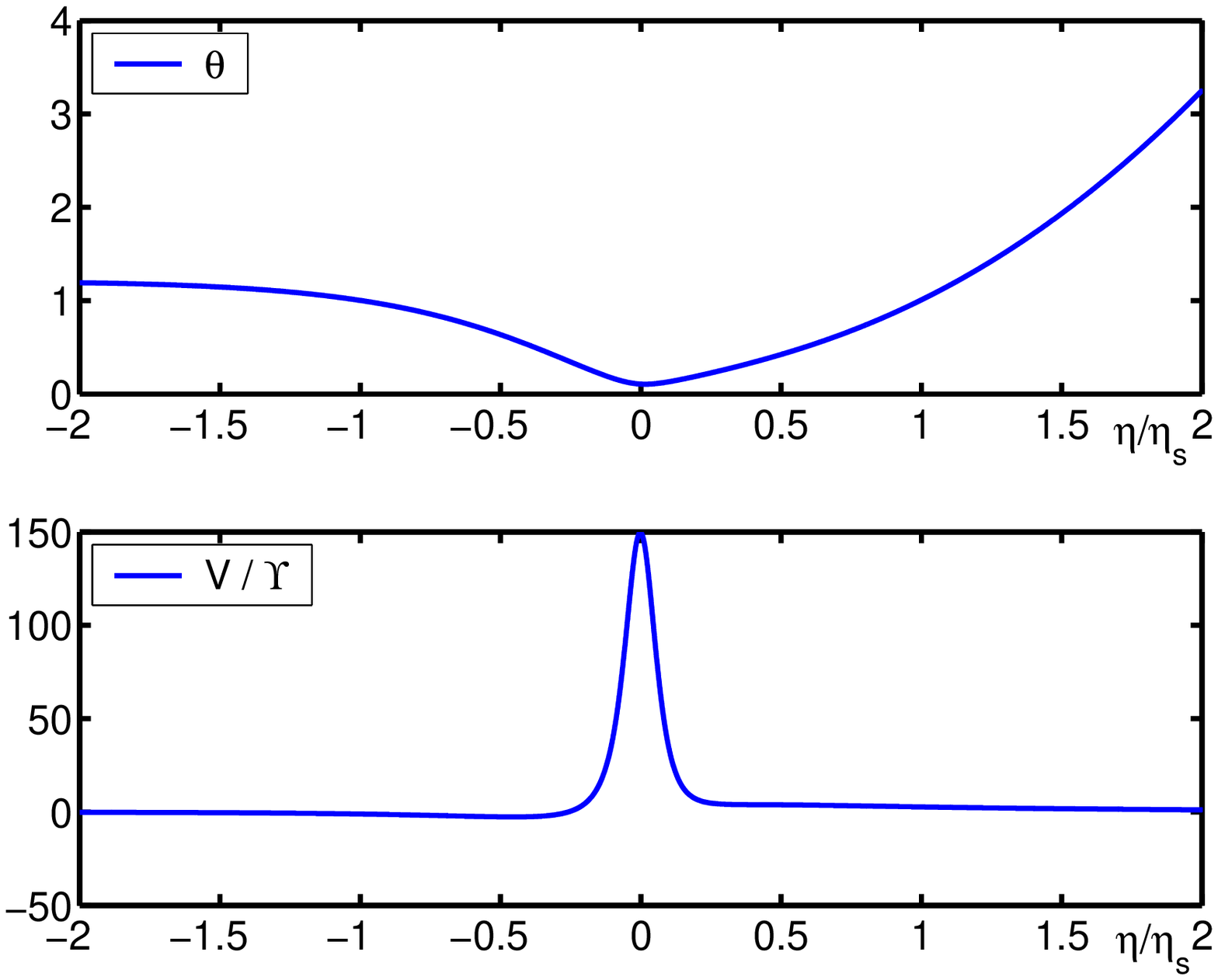}
 \quad
 \includegraphics[width=0.45\linewidth]{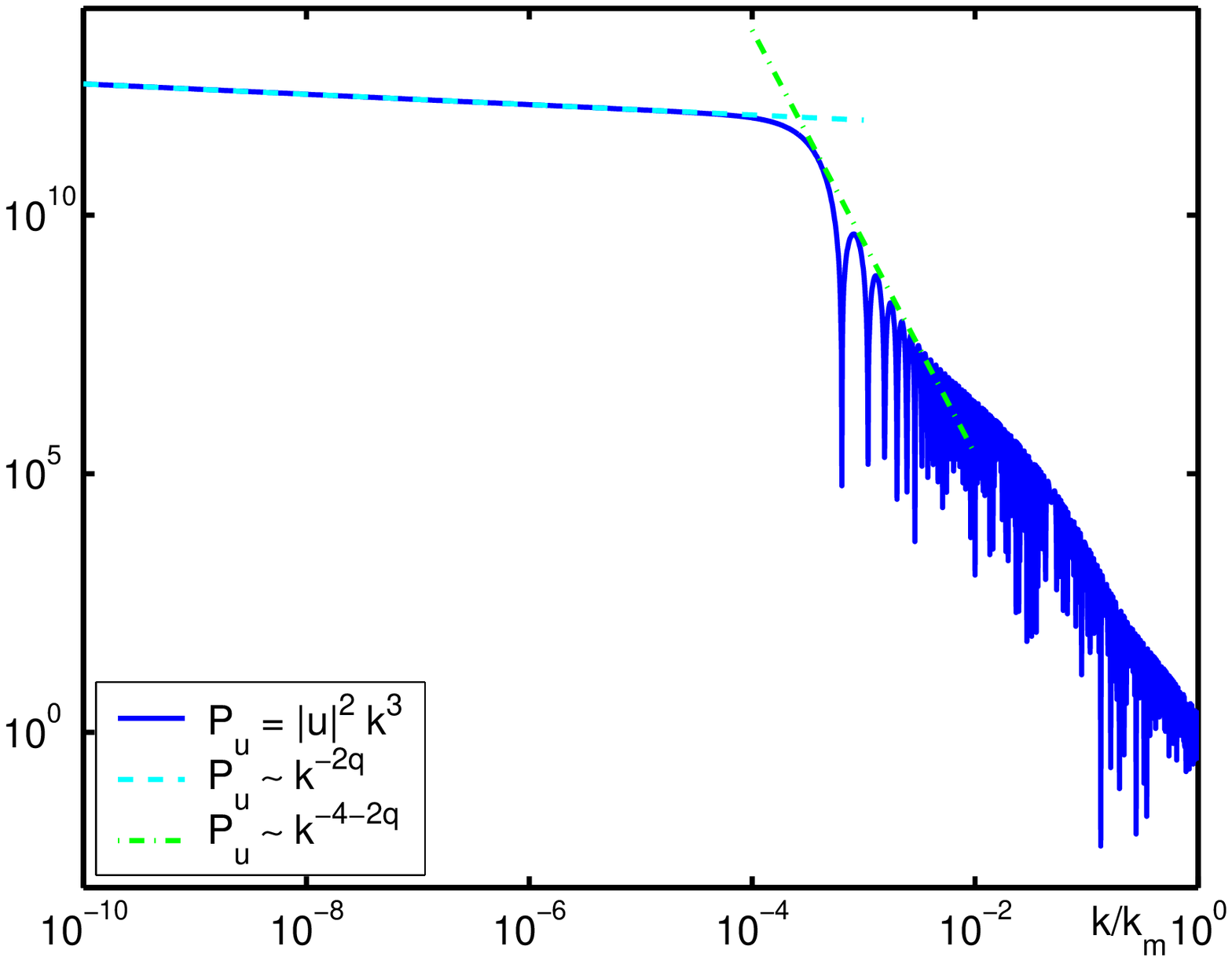}
 \caption
 {\label{f:case-2} \textbf{Case 2},
$\theta_1 \to \theta_2$, using $\tilde{\gamma}(\eta)
=-\iota(2+q)+2$. The $u$ potential (left) and spectrum,
$P_u=|u|^2k^3$ (right) are shown for  $q=5\cdot 10^{-2}$  and
$\epsilon=10^{-2}$.}
 \end{center}
\end{figure}

\begin{figure}[ht]
 \begin{center}
 \includegraphics[width=0.45\linewidth]{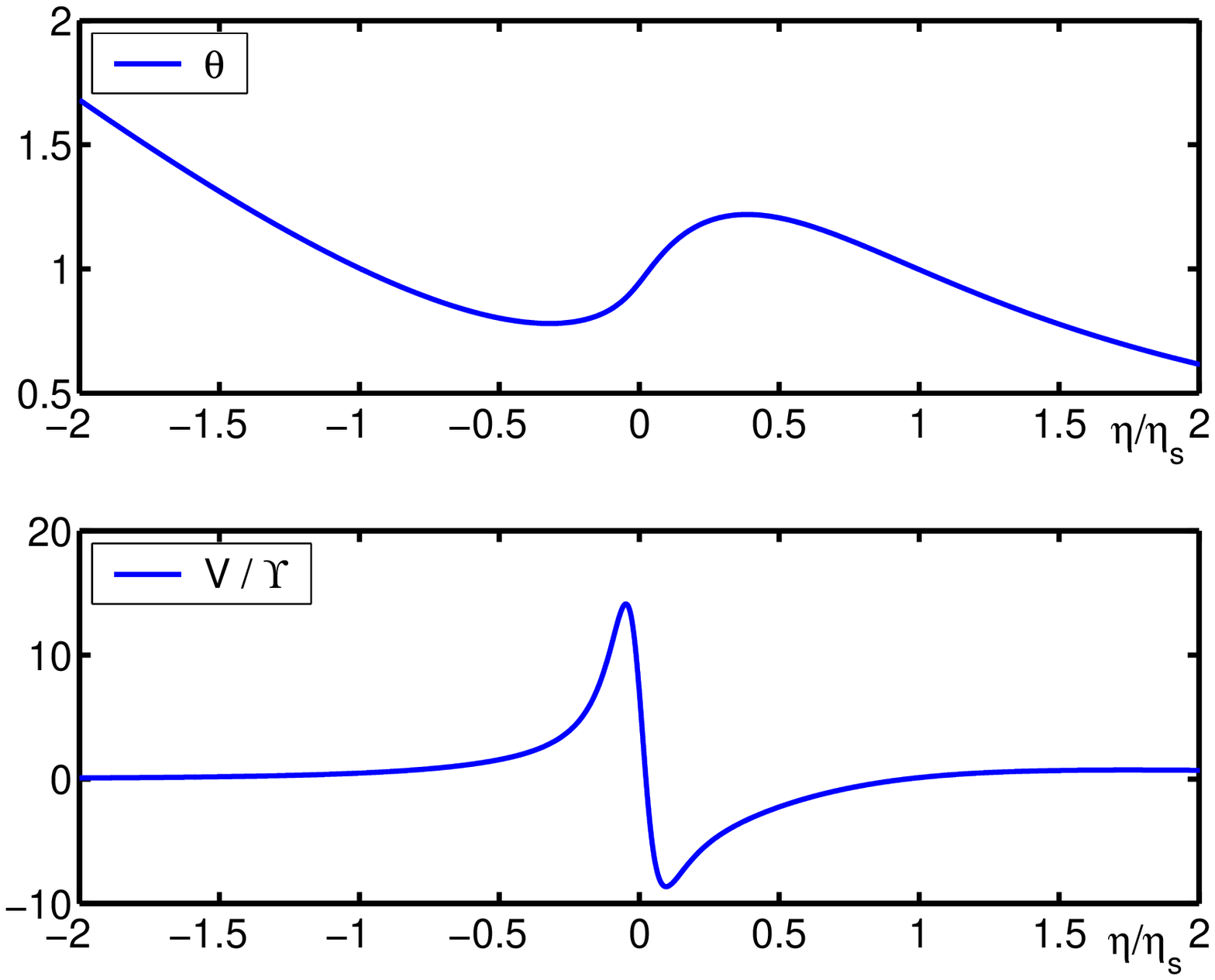}
 \quad
 \includegraphics[width=0.45\linewidth]{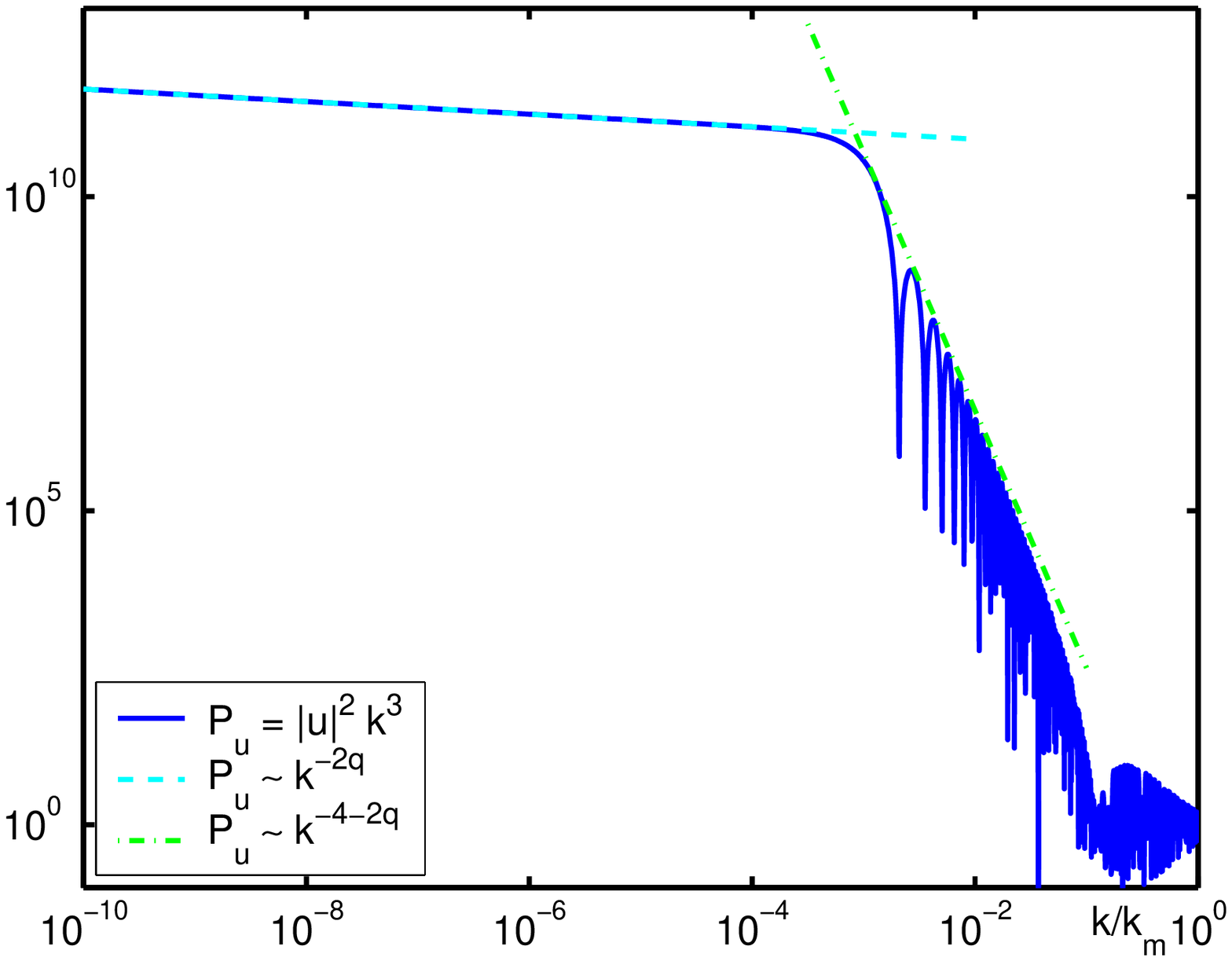}
 \caption
 {\label{f:case-3}  \textbf{Case 3}, $\theta_2 \to \theta_1$,
using $\tilde{\gamma}(\eta) =\iota(2+q)-1$. The $u$ potential
(left) and spectrum, $P_u=|u|^2k^3$ (right) are shown for
$q=5\cdot 10^{-2}$ and $\epsilon=10^{-2}$.}
 \end{center}
\end{figure}

\begin{figure}[ht]
 \begin{center}
 \includegraphics[width=0.45\linewidth]{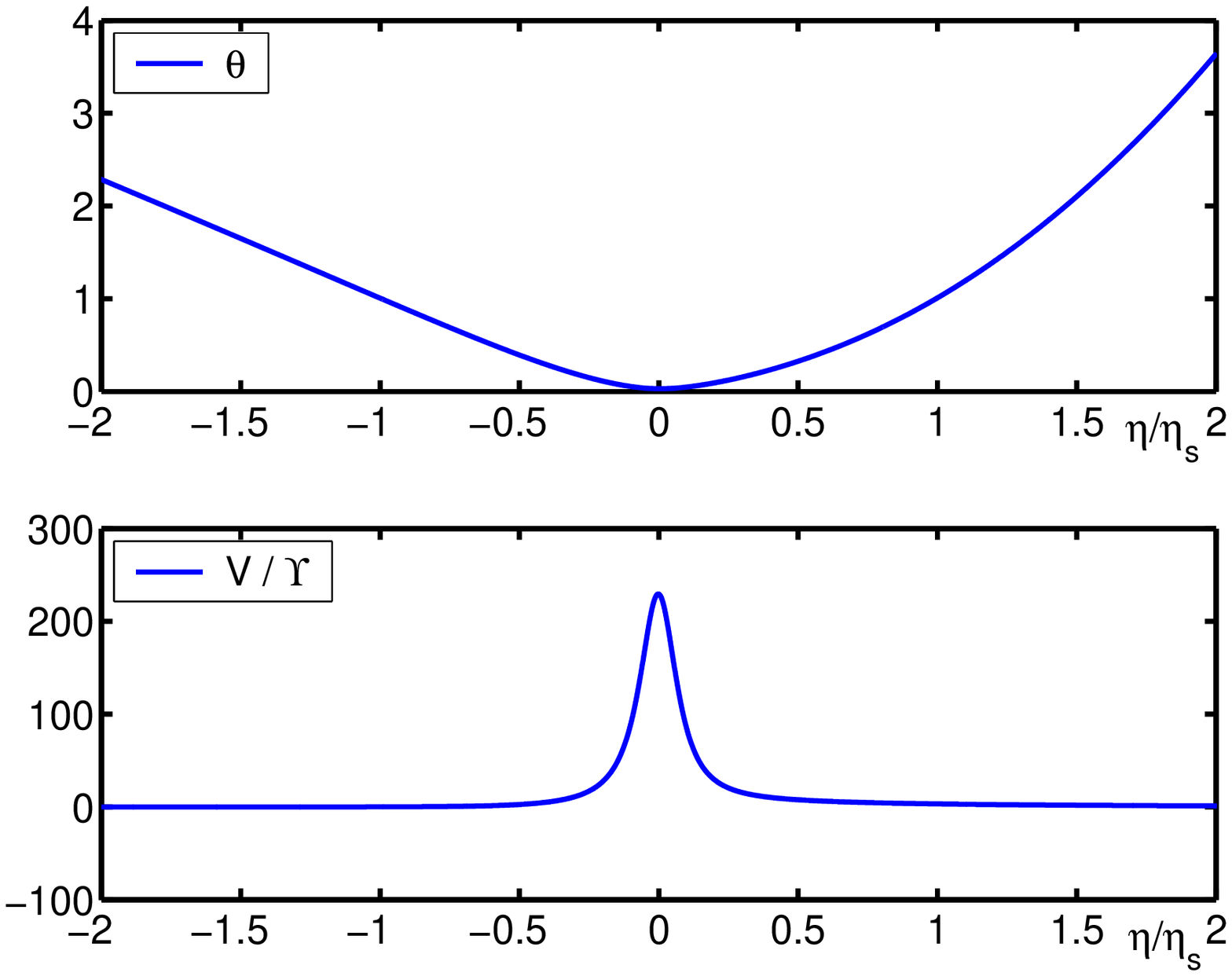}
 \quad
 \includegraphics[width=0.45\linewidth]{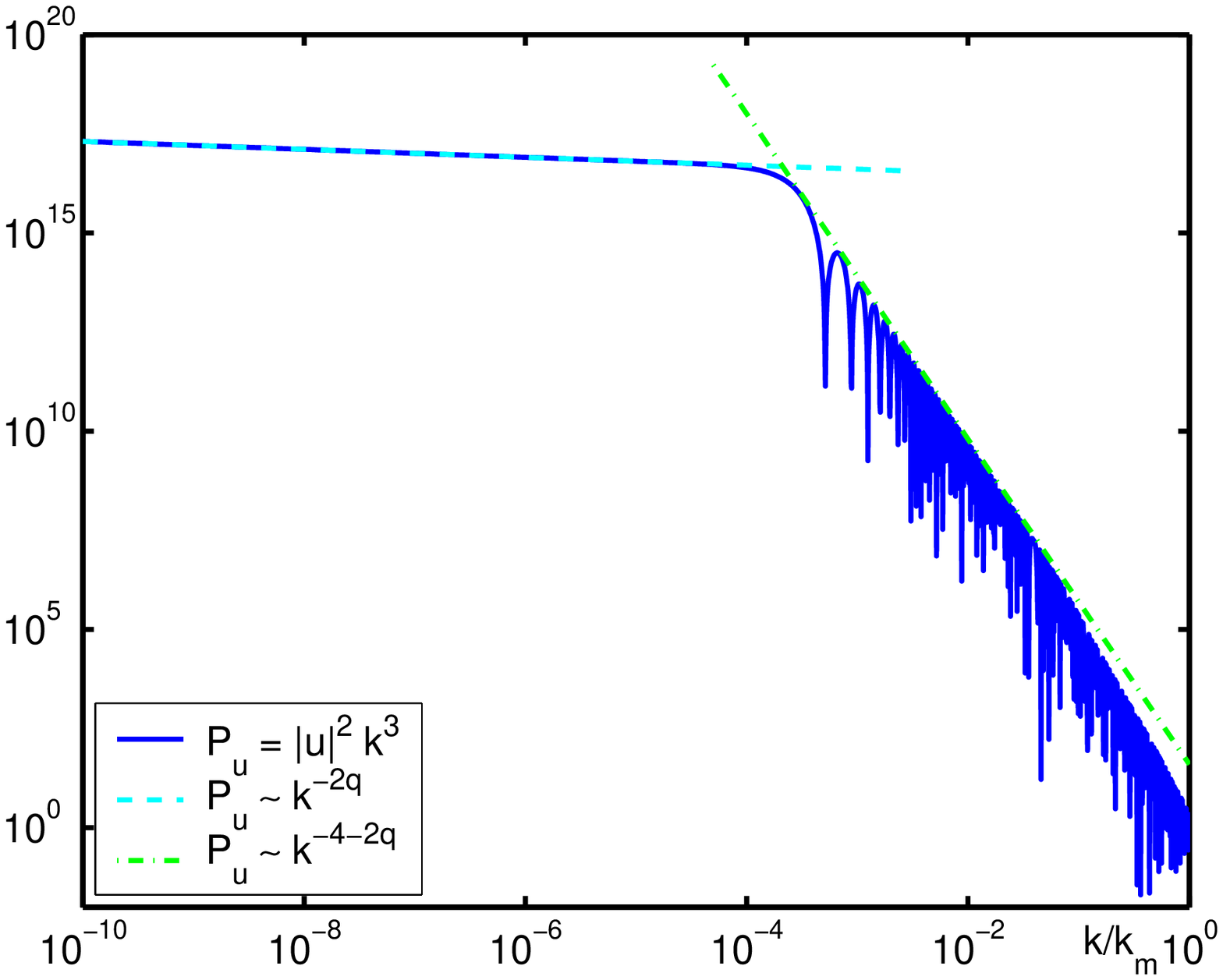}
 \caption
 {\label{f:case-4}  \textbf{Case 4},
$\theta_2 \to \theta_2$, using $\tilde{\gamma}(\eta) =
\iota(q-1)+2$. The $u$ potential (left) and spectrum,
$P_u=|u|^2k^3$ (right) are shown for $q=5\cdot 10^{-2}$ and
$\epsilon=10^{-2}$. }
 \end{center}
\end{figure}

\clearpage

It is very encouraging that the numerical simulations produce a
spectra with precisely the shape predicted analytically in the
previous section. The spectra are evaluated at $\eta=\eta_{end}
=10^3\eta_s$. As an example, we see that the kink predicted for
case~1 is there in the figure and is found at the correct
position,
\be
 k_u = \eta^{-1} \left({\eta\over\eta_s}\right)^{(2q-2)/(2q+1)}
 \simeq8\times 10^{-9}\eta_s^{-1} \simeq 10^{-9}k_m ~.
\ee
One also sees that the spectrum has the correct sub-Hubble radius
slope, $n_{sub}=n-4$ for scales $k>10^{-3}\eta_s^{-1}$ which have
already entered the Hubble radius at $\eta_{end}$.

It is interesting to note that the cases~2 and 3 have roughly the
same amplitude while the amplitude of case~4, which we expect to
depend on the transition, is much higher.

It is important to investigate the stability of the spectral index
$n=3+2q$ of case~1. To do this we have slightly modified the
potential for this case in the following way:
\begin{enumerate}
 \item For $\eta/\eta_s\in[-0.5,0.5]$,
   $V_u\to V_u +10^{-1}$, ~[(1) in Fig.~\ref{f:stability}].
 \item For $\eta/\eta_s\in[-0.5,0.5]$,
   $V_u\to V_u+10^{-3}$, ~[(2) in Fig.~\ref{f:stability}].
 \item For $\eta/\eta_s\in[-0.1,0.1]$,
   $V_u\to V_u+10^{-3}$, ~[(3) in Fig.~\ref{f:stability}].
\end{enumerate}

The result is shown in Fig~\ref{f:stability}. The plain curve
represents the original case, $V_u=\theta_1''/\theta_1$. Adding a
tiny constant $[\sim {\cal O}(10^{-3})]$ to this potential during
about a tenth of the duration of the transition [curve (3)]
already modifies significantly the amplification on super-Hubble
scales and the final spectral index becomes $n=1-2q$. Analyzing
the growing and decaying modes separately, we have seen that, due
to the perturbation of the potential, a tiny portion of the
growing mode during the pre-big-bang phase is converted into the
growing mode during the radiation era. This is already sufficient
for the latter to inherit the naively expected spectrum $n=1-2q$
like the other cases. The later we evaluate the spectrum the more
pronounced becomes the difference from the ``pure case~1''
spectrum. We expect that at very late times, hence very large
scales, extremely small differences from the pure case~1 potential
will have lead to a scale invariant spectrum. We have also tested
smooth modifications of the potential, like $V_u \to V_u +
10^{-3}\exp[-10^2(\eta/\eta_s)^2]$. They also lead to the same
result. The spectra of the cases~2 to 4, however, are stable under
small modifications of the corresponding potential.

\begin{figure}[ht]
 \begin{center}
 \includegraphics[width=0.45\linewidth]{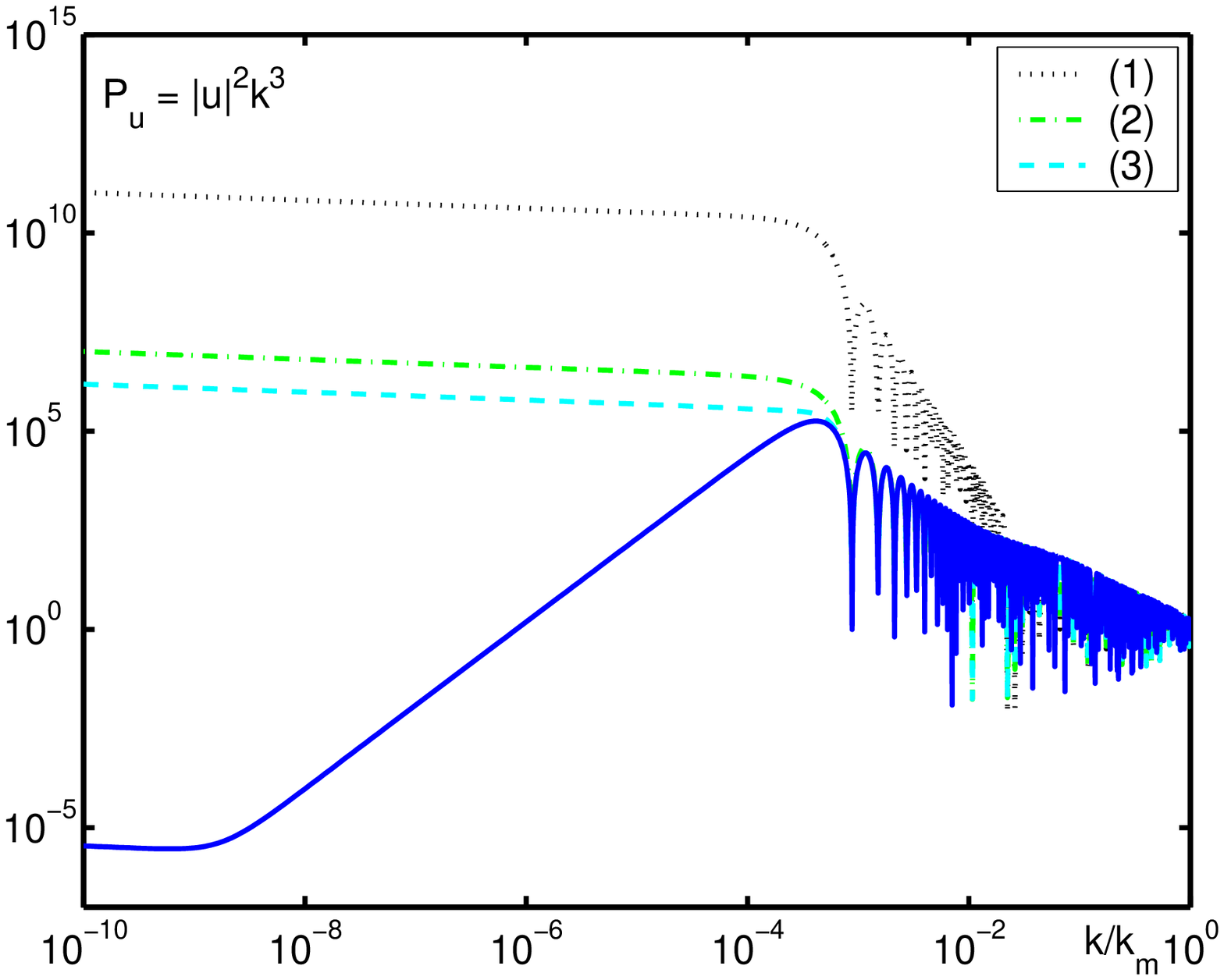}
 \quad
 \includegraphics[width=0.45\linewidth]{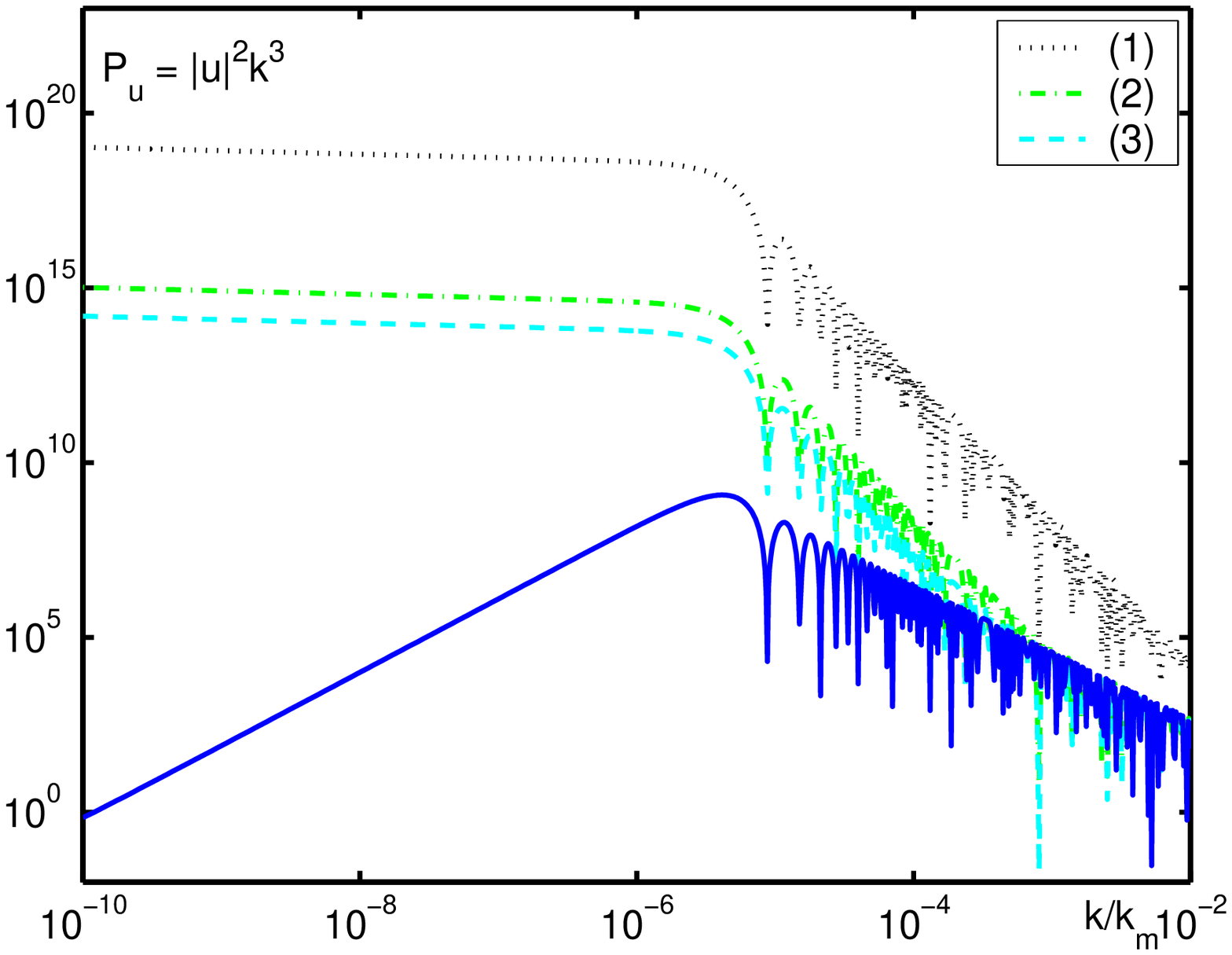}
 \caption
 {\label{f:stability}
We draw the spectral distribution ${\cal P}_{u} = |u|^2 k^3$
evaluated at $\eta/\eta_s = 10^3$ (left) and $\eta/\eta_s = 10^5$
(right) for $V_u$ of case 1 and the small modifications of the
potential detailed in the text. Although modifying the potential
during the transition may change $k_{m}$, we kept the same
original value of $k_{m}$ for all spectra.}
 \end{center}
\end{figure}

Finally, in Fig.~\ref{f:pbbu} we show the corresponding spectra
for dilaton-driven cosmology. The only difference to the previous
simulations is that we set $q=1/2$ in this model which, in the
Einstein frame, is a contracting universe with a scalar field with
vanishing potential. Again we obtain precisely the spectra
expected according to the arguments of the previous section.

\begin{figure}[ht]
 \begin{center}
 \includegraphics[width=0.45\linewidth]{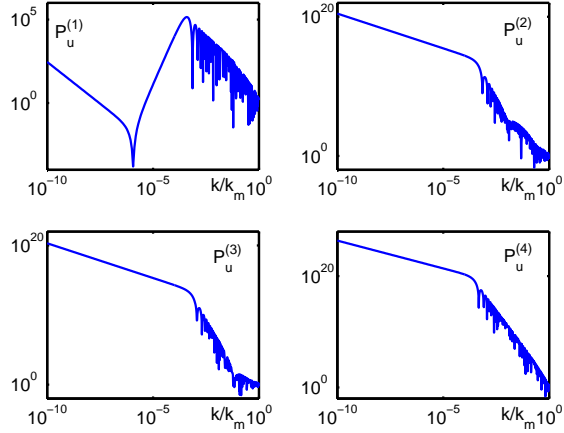}
\caption{We illustrate the spectral distribution ${\cal P}_u^{(i)}
= |u|^2 k^3$ (case $i=1-4$) as a function of the rescaled comoving
wave number $k/k_{m}$ The parameters for the simulations are
$\epsilon = 10^{-2}$, $q=0.5$ and the spectral distributions are
evaluated in the radiation era $\eta/\eta_s=10^3$.}
\label{f:pbbu}
 \end{center}
\end{figure}

\subsection{A regular transition in the perturbation variable $v$}

In this section we repeat the analysis presented above for the
case of a regular equation for the variable $v$. Since the
procedure is very close to the one presented above,  we can be
brief here. We again assume that there exists a regular potential
$V_v$ such that
\be
 v'' +(k^2\Upsilon-V_v)v = 0 ~.
\ee
In the case of a pure power law scale factor, $a=|\eta/\eta_s|^q$,
we have $V_v =q(q-1)/\eta^2 =z''/z$ where $z$ is either $z_1=a$ or
$z_2=a\int a^{-2}d\eta$. To regularize the $v$ equation during the
transition era we use our interpolation functions $\iota$,
$\tilde{\gamma}$ and $\Upsilon$ [see Eqs.~(\ref{e:def-iota}),
(\ref{e:def-gamma}) and (\ref{e:def-Upsilon}), respectively] and
impose
\be
 z(\eta)=\left[\left(\eta/\eta_s\right)^2
 +\epsilon\right]^{\tilde{\gamma}/2}~.
\ee
To reproduce the four cases for $v$, we make the following
choices:
\be \label{e:regpumpv}
 \tilde{\gamma}(\eta) = \left\{ \begin{array}{rccrcr}
  \iota(q-1)+ 1 & \mbox{ (case 1: } & z_1 \to z_1, &   q & \to & 1)~, \\
  \iota q       & \mbox{ (case 2: } & z_1 \to z_2, &   q & \to & 0)~, \\
 -\iota q +1    & \mbox{ (case 3: } & z_2 \to z_1, & 1-q & \to & 1)~, \\
  \iota(1-q)    & \mbox{ (case 4: } & z_2 \to z_2, & 1-q & \to &
  0)~.
\end{array} \right.
\ee
The resulting spectra are shown in Fig.~\ref{f:spectre-v-cases}.

\begin{figure}[ht]
 \begin{center}
 \includegraphics[width=0.45\linewidth]{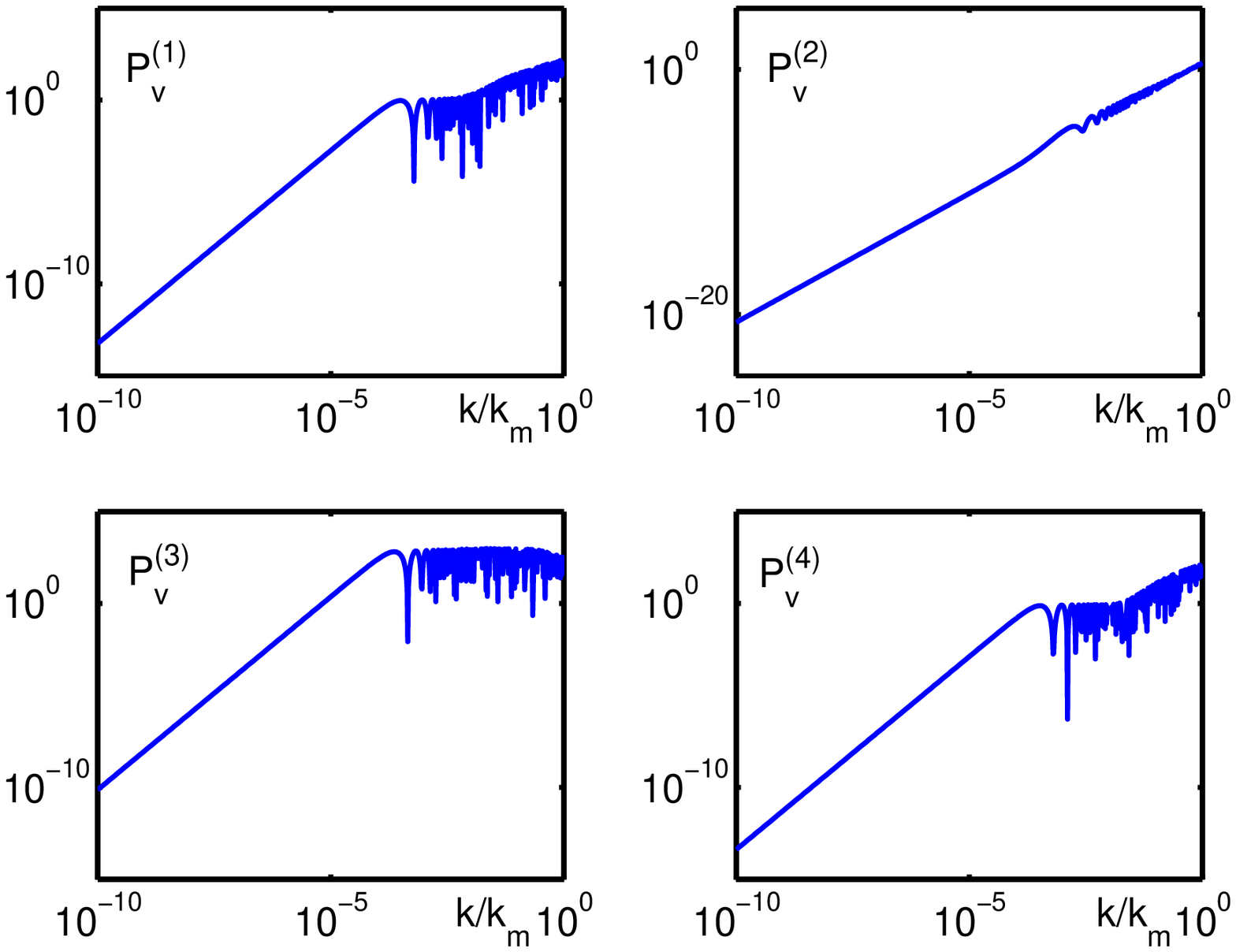}
\caption{We illustrate the spectral distribution ${\cal P}_v^{(i)}
= |v|^2 k^3$ (case $i=1-4$) as a function of the rescaled comoving
wave number $k/k_{m}$. The parameters for the simulations are
$\epsilon=10^{-2}$, $q=5\cdot10^{-2}$ and the spectral
distributions are evaluated in the radiation era at
$\eta/\eta_s=10^3$.}
 \label{f:spectre-v-cases}
 \end{center}
\end{figure}

Again the spectra are in very good agreement with those obtained
with our theoretical arguments. This gives us confidence that we
really understand what is going on. During the radiation era $v$
grows like $\eta$ on super-Hubble scales.  Inside the Hubble
radius the amplitude of $v$ remains constant and it begins to
oscillate. Neglecting the oscillations, we therefore expect for
the cases~1, 3 and 4
\be
 P_v \propto k^{2+2q}(\eta_s/\eta +k\eta)^{-2},
\ee
leading to the observed flat spectrum inside the Hubble radius.
The spectrum of case~2 turns from $P_v \propto k^{4-2q}$ to $P_v
\propto k^{2-2q}$ inside the Hubble radius. For $k\gsim 10^{-3}
k_m$ the spectrum is not very reliable, since it is influenced by
the details of the transition. The kink in the spectral
distribution expected for case $2$ is not well visible since
$q=5\cdot 10^{-2}$ is very small (see the argument on the previous
section). We have repeated this case for a larger value, $q=0.3$,
where the kink can now be seen, as illustrated in
Fig.~\ref{f:spectre-v-cases/fit-q=0.3}. Finally, note that case~2
for $v$ is unstable in the same sense as case~1 for $u$ and
therefore probably irrelevant.

\begin{figure}[ht]
 \begin{center}
 \includegraphics[width=0.45\linewidth]{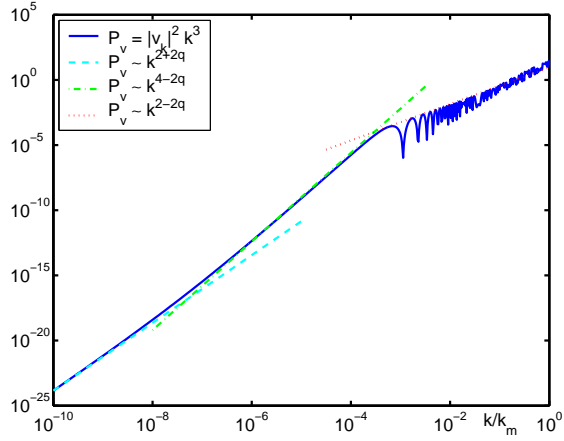}
 \caption{
We illustrate the spectral distribution ${\cal P}_v = |v|^2 k^3$
of case~2 as a function of the rescaled comoving wave number
$k/k_{m}$. The parameters for the simulations are $\epsilon =
10^{-2}$, $q=0.3$ and the spectral distribution is evaluated in
the radiation era at $\eta/\eta_s=10^3$. We confront our
predictions for the spectral slopes: above and below the kink at
$k_{v_1}/k_{m}\sim 10^{-8}-10^{-7}$, the decaying mode is still
dominant on scales $k\leqslant k_{v_1}$.}
 \label{f:spectre-v-cases/fit-q=0.3}
 \end{center}
\end{figure}

\section{Discussion}
\label{s:discussion}

\subsection{Which variable, $u$ or $v$?}

We have shown that during a contracting (or inflationary for
$q<0$) phase where the scale factor evolves according to a power
law, $a\propto |\eta|^q$ with $q>-1/2$, the variables $u$ and $v$
acquire a spectral index
\bea
 && P_u \propto k^{-2q}~, \quad~n_u =1-2q~, \\
 && P_v \propto k^{2+2q}~, \quad n_v =3+2q \quad
    \mbox{ if } q\leqslant1/2~, \\
 && P_v \propto k^{4-2q}~, \quad n_v =5-2q \quad
    \mbox{ if } q\geqslant 1/2~.
\eea
We have also shown that, if the corresponding variable transits
regularly into the radiation era, the spectrum is inherited in
this era. But, during the radiation era and on large scales
($k\eta\ll1$) $u$ and $v$ are simply related to $\Psi$, via [see
Eqs.~(\ref{e:def-u-Psi}), (\ref{e:def-zeta-Psi}) and
(\ref{e:def-v})]
\be
 \Psi =\left({M_s \over M_P}\right)\sqrt{2}a^{-2}u
      = -{\sqrt{6}\over 9}M_P^{-1}a^{-1}v~,
\ee
which therefore has the same spectrum as both $u$ and $v$. The
only possible resolution of this contradiction is that either $u$
or $v$ is not regular at the transition and makes a strongly
$k$-dependent jump. But which one?

We do not know the general answer to this question. It is quite
likely to be model dependent. Nevertheless, we are able to prove
the following statement.

\textit{Theorem.} If the perturbed metric remains regular during
the transition and if its evolution can be described by a second
order differential equation for the Bardeen potential $\Psi$, then
a regular $u$-variable which satisfies an equation of the form
(\ref{e:u-eqn}) can be found.

\textit{Proof.} If the metric perturbations are regular, the
Bardeen potential which is in general given by (see, e.g.
\cite{Durrer:1993db})
\be \label{e:genPsi}
 \Psi = A +k^{-1}{\cal H}(B-k^{-1}H_T') +k^{-1}(B-k^{-1}H_T')' ~,
\ee
is regular too. Here (scalar) metric perturbations are
parametrized with the four variables $A$, $B$, $H_L$ and $H_T$ via
\bea
ds^2 &=& \left[-(1+2A)d\eta^2 -i2B\hat k_jdx^jd\eta
     +\Big\{\Big(1+2H_L-\frac{2}{3}H_T\Big)\delta_{jm} \right. \nonumber \\
 && \quad \left.   +2 H_T\hat k_j\hat k_m\Big\}dx^jdx^m\right]
    a^2e^{-i{\bf k\cdot x}} ~,
\eea
and $\hat k_j =k_j/k$.

During a period governed by general relativity (and in the absence
of anisotropic stresses) the Bardeen potential satisfies an equation
of the form
\be
 \Psi'' +f(\eta)\Psi' +\left[\Upsilon(\eta) k^2 - g(\eta)\right]\Psi=0~,
\ee
where $f$, $g$ and $\Upsilon$ are smooth functions. If the scale
factor obeys a simple power law evolution, $a\propto |\eta|^q$, we
have $f=2(1+q)/\eta$ and $g=0$. If $f$ and $g$ remain smooth
during the transition, we can define
\be
 u = \exp\left({1\over 2}\int fd\eta\right)\Psi~.
\ee
As one easily verifies, this variable satisfies
\be
 u'' +(\Upsilon k^2 -V_u)u = 0
\ee
with
\be
 V_u = g +{1\over 2}f' + {1\over 4}f^2 ~.
\ee
$V_u$ is well defined, smooth and bounded in any finite interval,
so that the differential equation
\be
 \theta'' =V_u\theta
\ee
has two well-defined solutions $\theta_1$ and $\theta_2$ which are
the pump fields. Up to an irrelevant constant the so-defined
variable $u$ coincides with the well known $u$ given in
Eq.~(\ref{e:u-eqn}) in the asymptotic past $\eta\ll -\eta_s$ and
in the asymptotic future $\eta\gg \eta_s$. Hence it is our
regularized variable $u$. \hfill $\Box$

\medskip
On the other hand, if $\zeta$ passes via a regular second order
equation through the transition, the same theorem leads to a
regular $v$-equation and hence to a spectral index $n=3+2q$ for
$q\leqslant 1/2$.

This shows again that it is not possible for both $\Psi$ and
$\zeta$ to pass through the transition regularly (if $-1/2<q$).
This is  consistent with the expressions
Eq.~(\ref{e:def-zeta-Psi}) or Eq.~(\ref{e:zeta-constant}) which
relate $\zeta$  and $\Psi$. If these equations are also valid
during the transition,  $\zeta$ necessarily diverges if $\Psi$ is
regular and vice versa since ${\cal H}$ and ${\cal H}'-{\cal H}^2
= aH'$ have to pass through zero in a transition from contraction
to expansion (see also Fig.~\ref{f:short-transition}). Of course
these relations will in general be modified during the transition,
but according to our results the modifications should be such that
one of the two variables has to develop a singularity if the other
is regular.

\subsection{Amplitude of the perturbations}
\label{ss:amplitude}

During contraction the Bardeen potential grows like $\Psi \propto
|\eta|^{-(1+2q)}$ on super-Hubble scales. One actually has
\be
 |\Psi|^2k^3 \simeq \left({M_s\over M_P}\right)^2
  \left|\frac{\eta}{\eta_s}\right|^{-(2+4q)}(k\eta_s)^{-2q}
 \quad\mbox{ for } q\geqslant -1/2~.
\ee
Hence $\Psi$  may become much larger than $1$ for $k\ll 1/\eta_s$
and $|\eta| \sim \eta_s$. Does this imply that perturbation theory
breaks down during the contraction phase? We show now that this is
not the case for $q\leqslant 1$. First let us note that a quantity
relevant to measure the deviation of the geometry from Friedmann
is, for example, the Weyl curvature whose background component
vanishes. It is well known (see, e.g.~\cite{Durrer:1993db}) that
the ratio between a typical component of the Weyl tensor to a
typical component of the background Riemann tensor is given by
$|C/R| \simeq (k\eta)^2\Psi$. The geometrical deviation away from
Friedmann is thus of the order of
\be
 |C/R|^2k^3 \simeq  \left({M_s\over M_P}\right)^2
 \left|\frac{\eta}{\eta_s}\right|^{-4q}
    (k\eta_s)^{2-2q}(k\eta)^2~,
\ee
which is always much smaller than $1$ on super-Hubble scales for
$-1/2 \leqslant q\leqslant 1$ and $|\eta|\gsim \eta_s$. Only in a
contracting universe with $q>1$ do the perturbations become large
and hence perturbation theory becomes invalid.

To ensure that perturbations truly remain small, it is necessary
to find a gauge in which all the metric perturbations are small.
We show now that this is so in the  off-diagonal gauge which also
has been used in \cite{Brustein:1995kn} for dilaton-driven string
cosmology. This gauge is defined by $H_T=H_L=0$. According to
Eq.~(\ref{e:genPsi}), the Bardeen potential is then given by
\be
 \Psi= A +k^{-1}({\cal H} B + B') ~.
\ee
The $(ij)$ Einstein equation implies  (see e.g.
\cite{Durrer:1993db})
\be \label{e:PsiB}
 \Psi = A +k^{-1}({\cal H} B + B') = -k^{-1}{\cal H} B ~.
\ee
Before the transition, the Bardeen potential is given by
\be \Psi =
 {\sqrt{{\cal H}^2-{\cal H}'}\over M_Pa}u = \sqrt{q(q+1)}
 \left({M_s\over M_P}\right)
 \left|{\eta_s\over \eta}\right|^{1+q}u~.
\ee
On super-Hubble scales this gives for $q>-1/2$, using
Eq.~(\ref{e:u-superHubble}),
\bea
 |\Psi| &\simeq& \sqrt{q(q+1)}\left({M_s\over M_P}\right)
      k^{-3/2}(k\eta_s)^{-q}\left|{\eta_s\over \eta}\right|^{1+2q}
 \big[1 +{\cal O}(|k\eta|^\si) \big]\\
 \mbox{ with } && \si = \min(2q+1,2) . \nonumber
\eea
The exponent $\si > 0$ (for $q>-1/2$) of the first
correction to the dominant term can be obtained by expanding
the Hankel function solution given in
Eqs.~(\ref{e:u-initial-cond},\ref{e:def-u-Psi}).
With Eq.~(\ref{e:PsiB}) we then obtain
\bea  \label{e:B}
 B &\simeq & \sqrt{q+1\over q}\left({M_s\over M_P}\right)k^{-3/2}
             (k\eta_s)^{1-q} \left|{\eta_s\over \eta}\right|^{2q}
            \left[1 + {\cal O}\left(|k\eta|^\si\right)\right] ~, \\
 k^3|B|^2 & \simeq & \left({M_s\over M_P}\right)^2|k\eta|^{2-2q}
             \left|{\eta_s\over \eta}\right|^{2+2q} ~,
\eea
which remains small on super horizon scales, $|k\eta| <1$ during the
entire contraction phase, $\eta\leqslant -\eta_s$, for $q\leqslant 1$ .
From $A=-k^{-1}(2{\cal H} B + B')$ and Eq.~(\ref{e:B}) we find
that $ {\cal O}(A) = {\cal O}(|k\eta|^{\si-1} B)$, hence
\be
 k^3|A|^2  \simeq  \left({M_s\over M_P}\right)^2|k\eta|^{2\si-2q}
             \left|{\eta_s\over \eta}\right|^{2+2q} ~.
\ee
Inserting the value of $\si$ given above this becomes
\be
 k^3|A|^2  \simeq  \left\{ \begin{array}{ll}
\left({M_s\over M_P}\right)^2|k\eta|^{2+ 2q}
             \left|{\eta_s\over \eta}\right|^{2+2q} & \mbox{ for }
             -1/2<q\leqslant 1/2~,\\
\left({M_s\over M_P}\right)^2|k\eta|^{4- 2q}
             \left|{\eta_s\over \eta}\right|^{2+2q} & \mbox{ for }
             1/2 \leqslant q~.\\
  \end{array}\right.
\ee
With $B$ also $A$ remains small on super-Hubble scales as long as
$q\leqslant 1$ (actually $A$ remains small even for $1<q\leqslant
2$). In this treatment we have neglected the logarithm corrections
which are present for $q=1/2$. Note that it is highly non-trivial
that $A$ remains small. This is due to the fact that $B\propto
a^{-2}(1+{\cal O}|k\eta|^\si)$ and hence the lowest order
contribution to $A$ cancels.

Since the generic form of the perturbed Einstein equations is
${\cal O}[h +(k\eta)h +(k\eta)^2h] = {\cal O}(\Delta)$ where $h$
and $\Delta$ are typical metric and matter perturbation variables,
respectively (see, e.g. \cite{Durrer:1993db}), the matter
perturbation variables in this gauge are also small on
super-Hubble scales. More precisely we find from the perturbed
Einstein equations in the off-diagonal gauge
\bea  \label{e:de}
 \delta &=&  -{9\over 2}A + {3\over 2} (k/{\cal H})B  \simeq {\cal O}(A)~, \\
 \xi &=& {2k\over 3{\cal H}(1+w)}A +B  \simeq  B~, \\
 \pi_L &=& {2\over 3w}\left[{\cal H}^{-1}A' +{\cal H}^{-2}(2{\cal H}'+{\cal
     H}^2)A\right] \simeq {\cal O}(A)~. \label{e:piL}
\eea
Here $\delta$ and $\pi_L$ are the density and pressure
perturbations, respectively, and $i(\rho+P)\hat k_j\xi$ is the
scalar perturbation of the energy flux, $T^0_j$. To obtain the
above results we have used the fact that $a\propto |\eta|^q$ obeys
a simple power law and $B \propto |\eta|^{-2q}$ as well as $A
\propto |\eta|^{\max(0,1-2q)}$.

Our result that the perturbations remain small hold as long as
$\eta < -\eta_s$, which is the epoch when we expect higher order
curvature corrections to become important. Typically we would
expect this to be around the Planck scale. After that, what will
happen depends on the specific model considered.

%This result seems in contradiction with the statement of
%Ref.~\cite{Lyth:2001nv} that perturbations in the 'new' ekpyrotic
%universe, $q= 1/2$ necessarily become large for $\eta\to 0$. The
%contradiction is quickly revealed: In Ref.~\cite{Lyth:2001nv} the
%author considers the curvature variable $\cal R$ which corresponds
%to our $A$ up to a factor of order unity. Not neglecting the
%$\log$-term, he correctly finds that this variable diverges for
%$\eta\ra 0$. But as we have argued above, this solution is only to
%be taken seriously for $\eta < -\eta_s$ where it remains small,
%for all values of $\eta_s$ hence also in the limit $\eta_s\ra 0$.
%The difference of that conclusion to Ref.~\cite{Lyth:2001nv}, is
%the way in which we perform the limit $\eta \ra 0$. We perform it
%by requiring $\eta < -\eta_s$ be always satisfied, hence $\eta \ra
%0$ requires that also  $\eta_s \ra 0$, while in
%Ref.~\cite{Lyth:2001nv}, the normalization constant  $\eta_s$
%(which is not explicitly introduced) is kept fixed.

\section{Conclusion}
In this work we have analyzed the behavior of scalar perturbations
in a transition from a contracting to an expanding Friedmann
universe. We have shown that, if the perturbation equation during
the transition can be formulated as second order equations for
either $\Psi$ or $\zeta$, regular variables $u$ or $v$,
respectively, can be found. The resulting spectral index in the
late radiation dominated universe {\em depends} on which of these
two variables passes regularly, and there are no stable cases
where both $u$ and $v$ (equivalently $\Psi$ and $\zeta$), are
regular during the transition.

The resulting spectral index $n$ is given by
\be \label{e:specfin}
 n = \left\{ \begin{array}{ll}
     1-2q & \mbox{if $\Psi$ is regular}, \\
     3+2q & \mbox{if $\zeta$ is regular and $q\leqslant 1/2$}, \\
     5-2q & \mbox{if $\zeta$ is regular and $q > 1/2$}.
     \end{array}\right.
\ee
Our numerical results for the spectral index obtained from a
simple toy model are in perfect agreement with the more general
arguments of Sec.~\ref{s:General solutions}.

This result remains valid in an inflationary universe with
$-1/2<q<0$, but has never raised any attention since such models
cannot produce the observed scale invariant spectrum and require
$w<-1$. For $q<-1/2$ both variables $u$ and $v$ lead to the same
spectral index $n=3+2q$. Therefore, this problem has not been
noticed in works on standard inflationary models where $q\simeq
-1$.

We have also shown that, as long as $q\leqslant 1$, and we are in
a regime where corrections to the equations of motion can be
ignored, perturbations remain small during contraction in the
sense that there exists a gauge in which all the metric and matter
perturbation variables are small.

We have also argued that the $v$ equation derived from string
corrections in~\cite{Cartier:2001is} has to be considered as a toy
model, since higher order corrections cannot be neglected in this
case.

Our findings explain that all the literature based on the variable
$v$ predicts $n=3+2q$, see especially Refs.~\cite{Cartier:2001is}
and \cite{Tsujikawa:2002qc}, while when mainly working with $u$
one finds that the spectral index $n=3+2q$ is highly unstable and
one typically expects $n=1-2q$.

This work has the following important implications.
\begin{itemize}
\item  {\em If} it can be shown that in the ekpyrotic model
\cite{Khoury:2001wf,Khoury:2001zk} where $0<q\ll 1$ the
Bardeen potential passes regularly through the transition, this
model leads to a nearly scale invariant spectrum with $n=1-2q$.

\item In dilaton-driven string cosmology we have the opposite
situation. There, $q=1/2$ and it has generically been assumed that
$\zeta$ passes regularly through the transition. This has been
shown to be true to first order in $\alpha'$
in~\cite{Cartier:2001is}. Then the spectral index is $n=3+2q=4$
leading to a very blue spectrum of highly suppressed
perturbation~\cite{Brustein:1995kn}. In this case, a scale
invariant spectrum of adiabatic perturbations of the axion field
can be obtained via the ``curvaton mechanism''
\cite{Bozza:2002fp}. If, however, $\Psi$ would be regular, a red
spectrum with $n=1-2q$ is obtained. This would mean a fatal blow
for dilaton-driven string cosmology, since the perturbations then
become very large in the radiation dominated era: Since the
Bardeen potential is large at the end of the pre-big-bang phase
and since $ |C/R| \simeq (k\eta)^2|\Psi| \simeq |\Psi|$ at Hubble
crossing, the Weyl tensor becomes larger than the background
Riemann tensor at Hubble crossing. Even though the Weyl tensor has
constant amplitude on super-Hubble scales, the decay of the
Riemann tensor during expansion leads to a huge increase in the
ratio $ |C/R|$. This problem only affects red spectra, since for
blue spectra $ |C/R|$ at Hubble crossing is always smaller than
$|\Psi(k_{\max})| \simeq |C/R|$ evaluated at $\eta=-\eta_s$ for
$k=k_{\max}$.
\end{itemize}

Even though we cannot establish from first principles in this work
which spectrum dilaton-driven string cosmology or the ekpyrotic
model have, we nevertheless have formulated sufficient (but maybe
not necessary) conditions on the transition which would allow such
a decision.

\section*{Acknowledgments}
 We thank Robert Brandenberger, Patrick
Peter, Gabriele Veneziano, Filippo Vernizzi and David Wands for
stimulating and clarifying discussions. We are grateful to David
Lyth who helped us to correct an erroneous equation in the first
draft and for additional stimulating correspondence. C.C.
acknowledges financial support from the Tomalla Foundation. E.C.
and R.D. thank the Aspen Center of Physics for hospitality. This
work is supported by the Swiss National Science Foundation.

%======================================%
%<<<<<<<<<< END MAIN TEXT >>>>>>>>>>>>>%
%======================================%

%\newpage
%\bibliographystyle{../../Bibstyle/xxx-prd}
%\bibliography{../../bib-art,../../bib-book,../../bib-temp}

\begin{thebibliography}{10}

\bibitem{Khoury:2001zk}
J.~Khoury, B.~A. Ovrut, P.~J. Steinhardt, and N.~Turok,
\newblock Phys. Rev. D {\bf 66}, 046005 (2002),
  [\href{http://xxx.lanl.gov/abs/hep-th/0109050}{hep-th/0109050}].

\bibitem{Durrer:2002jn}
R.~Durrer and F.~Vernizzi,
\newblock Phys. Rev. D {\bf 66}, 083503 (2002),
  [\href{http://xxx.lanl.gov/abs/hep-ph/0203275}{hep-ph/0203275}].

\bibitem{Peter:2002cn}
P.~Peter and N.~Pinto-Neto,
\newblock Phys. Rev. D {\bf 66}, 063509 (2002),
  [\href{http://xxx.lanl.gov/abs/hep-th/0203013}{hep-th/0203013}].

\bibitem{Khoury:2001wf}
J.~Khoury, B.~A. Ovrut, P.~J. Steinhardt, and N.~Turok,
\newblock Phys. Rev. D {\bf 64}, 123522 (2001),
  [\href{http://xxx.lanl.gov/abs/hep-th/0103239}{hep-th/0103239}].

\bibitem{Brandenberger:2001bs}
R.~Brandenberger and F.~Finelli,
\newblock JHEP {\bf 11}, 056 (2001),
  [\href{http://xxx.lanl.gov/abs/hep-th/0109004}{hep-th/0109004}].

\bibitem{Lyth:2001pf}
D.~H. Lyth,
\newblock Phys. Lett. B {\bf 524}, 1 (2002),
  [\href{http://xxx.lanl.gov/abs/hep-ph/0106153}{hep-ph/0106153}].

\bibitem{Hwang:2001ga}
J.-c. Hwang,
\newblock Phys. Rev. D {\bf 65}, 063514 (2002),
  [\href{http://xxx.lanl.gov/abs/astro-ph/0109045}{astro-ph/0109045}].

\bibitem{Gratton:2003pe}
S.~Gratton, J.~Khoury, P.~J. Steinhardt, and N.~Turok,
\newblock (2003),
  [\href{http://xxx.lanl.gov/abs/astro-ph/0301395}{astro-ph/0301395}].

\bibitem{Antoniadis:1994jc}
I.~Antoniadis, J.~Rizos, and K.~Tamvakis,
\newblock Nucl. Phys. {\bf B415}, 497 (1994),
  [\href{http://xxx.lanl.gov/abs/hep-th/9305025}{hep-th/9305025}].

\bibitem{Brustein:1998cv}
R.~Brustein and R.~Madden,
\newblock Phys. Rev. D {\bf 57}, 712 (1998),
  [\href{http://xxx.lanl.gov/abs/hep-th/9708046}{hep-th/9708046}].

\bibitem{Foffa:1999dv}
S.~Foffa, M.~Maggiore, and R.~Sturani,
\newblock Nucl. Phys. {\bf B552}, 395 (1999),
  [\href{http://xxx.lanl.gov/abs/hep-th/9903008}{hep-th/9903008}].

\bibitem{Cartier:1999vk}
C.~Cartier, E.~J. Copeland, and R.~Madden,
\newblock JHEP {\bf 01}, 035 (2000),
  [\href{http://xxx.lanl.gov/abs/hep-th/9910169}{hep-th/9910169}].

\bibitem{Cartier:2001is}
C.~Cartier, J.~C. Hwang, and E.~J. Copeland,
\newblock Phys. Rev. D {\bf 64}, 103504 (2001),
  [\href{http://xxx.lanl.gov/abs/astro-ph/0106197}{astro-ph/0106197}].

\bibitem{Tsujikawa:2002qc}
S.~Tsujikawa, R.~Brandenberger, and F.~Finelli,
\newblock Phys. Rev. D {\bf 66}, 083513 (2002),
  [\href{http://xxx.lanl.gov/abs/hep-th/0207228}{hep-th/0207228}].

\bibitem{Finelli:2001sr}
F.~Finelli and R.~Brandenberger,
\newblock Phys. Rev. D {\bf 65}, 103522 (2002),
  [\href{http://xxx.lanl.gov/abs/hep-th/0112249}{hep-th/0112249}].

\bibitem{Mukhanov:1992tc}
V.~F. Mukhanov, H.~A. Feldman, and R.~H. Brandenberger,
\newblock Phys. Rept. {\bf 215}, 203 (1992).

\bibitem{Durrer:1993db}
R.~Durrer,
\newblock Fund. Cos. Phys. {\bf 15}, 209 (1994),
  [\href{http://xxx.lanl.gov/abs/astro-ph/9311041}{astro-ph/9311041}].

\bibitem{Lyth:1985aa}
D.~H. Lyth,
\newblock Phys. Rev. D {\bf 31}, 1792 (1985).

\bibitem{Veneziano:1991ek}
G.~Veneziano,
\newblock Phys. Lett. B {\bf 265}, 287 (1991),
  [\href{http://xxx.lanl.gov/abs/CERN-TH.6077/91}{CERN-TH.6077/91}].

\bibitem{Gasperini:2002bn}
M.~Gasperini and G.~Veneziano,
\newblock Phys. Rep. {\bf 373}, 1 (2003),
  [\href{http://xxx.lanl.gov/abs/hep-th/0207130}{hep-th/0207130}].

\bibitem{Brustein:1995kn}
R.~Brustein, M.~Gasperini, M.~Giovannini, V.~F. Mukhanov, and
G.~Veneziano,
\newblock Phys. Rev. D {\bf 51}, 6744 (1995),
  [\href{http://xxx.lanl.gov/abs/hep-th/9501066}{hep-th/9501066}].

\bibitem{Deruelle:1995kd}
N.~Deruelle and V.~F. Mukhanov,
\newblock Phys. Rev. D {\bf 52}, 5549 (1995),
  [\href{http://xxx.lanl.gov/abs/gr-qc/9503050}{gr-qc/9503050}].

\bibitem{Page:1984}
D.~N. Page,
\newblock Class. Quantum Grav. {\bf 1}, 417 (1984).

\bibitem{Wands:1998yp}
D.~Wands,
\newblock Phys. Rev. D {\bf 60}, 023507 (1999),
  [\href{http://xxx.lanl.gov/abs/gr-qc/9809062}{gr-qc/9809062}].

\bibitem{Brustein:1998kq}
R.~Brustein, M.~Gasperini, and G.~Veneziano,
\newblock Phys. Lett. B {\bf 431}, 277 (1998),
  [\href{http://xxx.lanl.gov/abs/hep-th/9803018}{hep-th/9803018}].

\bibitem{Heard:2002dr}
I.~P.~C. Heard and D.~Wands,
\newblock Class. Quant. Grav. {\bf 19}, 5435 (2002),
  [\href{http://xxx.lanl.gov/abs/gr-qc/0206085}{gr-qc/0206085}].

\bibitem{Bozza:2002fp}
V.~Bozza, M.~Gasperini, M.~Giovannini, and G.~Veneziano, Phys.
Lett. B {\bf 543}, 14 (2002),
  [\href{http://xxx.lanl.gov/abs/hep-ph/0206131}{hep-ph/0206131}].

\end{thebibliography}
%\addcontentsline{toc}{section}{References}
%\end{document}

\providecommand{\href}[2]{#2}\begingroup\raggedright

\end{document}